\documentclass[a4paper,11pt]{article}
\pdfoutput=1 

\usepackage{jheppub} 

\usepackage[T1]{fontenc} 
\usepackage{placeins}
\usepackage[dvipsnames,svgnames,x11names,hyperref]{xcolor}
\usepackage{array}
\usepackage{graphicx}
\usepackage{booktabs}
\usepackage{slashed}
\RequirePackage{mathrsfs}
\usepackage{dsfont}
\usepackage{hyperref}
\hypersetup{
        unicode = true,
        pdftitle = Note , 
        colorlinks = true,
        linkcolor = Black!30!DodgerBlue4,
        citecolor = Black!30!DodgerBlue4,
        filecolor = Black!30!DodgerBlue4,
        urlcolor = Black!30!DodgerBlue4,
}

\usepackage{booktabs}

\renewcommand{\toprule}{\specialrule{1.5pt}{0em}{1pt} \midrule}
\renewcommand{\bottomrule}{\midrule \specialrule{1.5pt}{1pt}{0em}}

\graphicspath{{./plots/}}

\newcommand{\erf}{\mathop{\mathrm{erf}}}

\newcommand{\hc}{\mathrm{h.c.}}

\renewcommand{\epsilon}{\varepsilon}

\renewcommand{\vec}[1]{\boldsymbol{#1}}
\renewcommand{\bar}{\overline}
\renewcommand{\Re}{\mathrm{Re}}
\renewcommand{\Im}{\mathrm{Im}}

\newcommand{\ct}[1]{c^\tau_{#1}}
\newcommand{\ctp}[1]{c^{\tau'}_{#1}}
\newcommand{\sX}{\vec{S}_{X}^{s' s}}
\newcommand{\sn}{\vec{S}_{N}^{r' r}}
\newcommand{\sXno}{\vec{S}_{X}}
\newcommand{\snno}{\vec{S}_{N}}
\newcommand{\sy}{\boldsymbol{\mathcal{S}}}
\newcommand{\Op}{\mathcal{O}}
\newcommand{\vt}{\vec{v}^{\bot}}
\newcommand{\qmn}{\frac{\vec{q}}{m_N}}
\newcommand{\qtmn}{\frac{\vec{q}}{2m_N}}
\newcommand{\tqmn}{\frac{2\vec{q}}{m_N}}
\newcommand{\qpmn}[1]{\frac{\vec{q}\,^{#1}}{m_N^{#1}}}

\newcolumntype{L}{>{$}l<{$}}
\newcolumntype{C}{>{$}c<{$}}
\newcommand{\bra}[1]{\left\langle #1\right|}
\newcommand{\ket}[1]{\left|#1\right\rangle}
\newcommand{\es}{\vec{e}_s}
\newcommand{\esp}{\vec{e}\,'_{s'}}
\allowdisplaybreaks

\title{\boldmath Non-relativistic Effective Interactions of Spin 1 Dark Matter}

\author[a]{Riccardo Catena}
\author[a,b]{K\aa re Fridell}
\author[a,c]{and Martin B. Krauss}

\affiliation[a]{Chalmers University of Technology, Department of Physics, SE-412 96 G\"oteborg, Sweden}
\affiliation[b]{Fakult\"at f\"ur Physik, Technische Universit\"at M\"unchen, James-Franck-Stra\ss e 1/II, 85748 Garching, Germany}
\affiliation[c]{Dipartimento di Matematica e Fisica, Università di Roma Tre, Via della Vasca Navale 84, 00146 Rome, Italy}

\emailAdd{catena@chalmers.se}
\emailAdd{kare.fridell@tum.de}
\emailAdd{martin.krauss@chalmers.se}

\abstract{We investigate the non-relativistic reduction of simplified models for spin 1 dark matter (DM) with the aim of identifying features in the phenomenology of DM-quark interactions which are specific to vector DM.~In the case of DM-quark interactions mediated by a spin 1 particle, we find two DM-nucleon interaction operators arising from the non-relativistic reduction of simplified models for spin 1 DM that are specific to spin 1 DM, and which were not considered in previous studies.~They are quadratic in the momentum transfer, linear in a symmetric combination of polarisation vectors for the DM particle, and arise from simplified models which do not generate momentum transfer independent operators as leading interactions in the non-relativistic expansion of DM-nucleon scattering amplitudes.~Within these simplified models, the new operators cannot be neglected when computing DM signals at direct detection experiments.~For example, we find that nuclear recoil energy spectra computed by including or neglecting the new operators can differ by up to one order of magnitude for nuclear recoil energies larger than about 20 keV and DM masses below $50$~GeV.~Furthermore, the shape of the expected nuclear recoil spectra depends significantly on whether the new operators are taken into account or not.~Finally, neglecting the contribution to DM direct detection signals from the new operators leads to inaccurate conclusions when assessing the compatibility of a future direct detection signal with CMB constraints on the DM relic density, especially when the number of signal events is small, e.g.~$\mathcal{O}(1)$.} 

\begin{document} 
\maketitle
\flushbottom

\section{Introduction}
Measurements of velocity dispersion in stellar and galactic systems, gravitational lensing events in galaxy clusters, the hierarchical formation of large scale cosmological structures, and the second peak in the angular power spectrum of the cosmic microwave background radiation show that invisible mass, or Dark Matter (DM), must be present in our Universe~\cite{Bertone:2016nfn}.~While the microscopic constituents of DM have so far escaped detection, complementary experimental methods are currently used to search for the elusive ``DM particle''~\cite{Bertone:2004pz}.~Direct detection experiments primarily search for nuclear recoils induced by the non-relativistic scattering of Milky Way DM particles in low-background experiments located deep underground~\cite{Drukier:1983gj,Goodman:1984dc}.~For a review of current direct detection methods, see for example~\cite{Undagoitia:2015gya}.

Predictions for the expected rate of nuclear recoil events at direct detection experiments have been refined significantly in recent years, e.g.~\cite{DelNobile:2018dfg}.~Special emphasis has been placed on modelling the ``elementary'' DM-nucleon interactions and computing the associated nuclear response functions~\cite{Anand:2013yka,Vietze:2014vsa,Hoferichter2015,Hoferichter:2018acd}.~Three approaches have been pursued to model the interactions of DM with nucleons:~1) non-relativistic effective theories for DM-nucleon interactions~\cite{Fan:2010gt,Fitzpatrick:2012ix};~2) relativistic effective field theories~\cite{Brod:2017bsw}; 3) and, finally, simplified models~\cite{Abdallah:2015ter,DeSimone:2016fbz,Baum:2018sxd,Baum:2018lua}.~In non-relativistic effective theories the relevant degrees of freedom are DM and nucleons.~Their interactions are constrained by Galilean invariance and built in terms of a set of basic non-relativistic quantum mechanical operators (see Eq.~(\ref{eq:list_of_small_ops})).~Relativistic effective field theories for DM direct detection are built in terms of higher dimensional DM-quark and -gluon interaction operators constrained by Lorentz invariance.~Their matching onto nucleon-level non-relativistic operators has been performed by using chiral effective field theory~\cite{Hoferichter:2015ipa,Bishara:2016hek}.~Not all operators appearing in non-relativistic effective theories for DM-nucleon interaction arise as leading operators through this matching procedure.~Finally, simplified models for DM extends the Standard Model of particle physics by one DM candidate and one additional particle that is responsible for the interactions of DM with quarks and gluons.~For a discussion on the constraints unitarity and anomaly cancellation impose on this latter approach to DM model building, see for example~\cite{Kahlhoefer:2015bea,Ellis:2017tkh,Ellis:2018xal}.~The three approaches are related as follows. When the momentum transferred in DM-nucleus interactions is smaller than the mass of the particle mediator, simplified models reduce to relativistic effective field theories.~When the momentum transferred is smaller than the nucleon mass and DM moves at non-relativistic speeds, each relativistic DM-quark and -gluon operator reduces to a linear combination of non-relativistic quantum mechanical operators for DM-nucleon interactions.~In the case of spin 0 and spin 1/2 DM, the relation between simplified models and non-relativistic quantum mechanical operators for DM-nucleon interactions has been investigated extensively in recent years~\cite{Crivellin:2014qxa,DEramo:2014nmf,DEramo:2016gos}.~On the other hand, the non-relativistic reduction of simplified models for spin 1 DM has been  significantly less explored.

This work focuses on the non-relativistic reduction of simplified models for spin 1 DM, exploring the cases of a spin 0 and spin 1 mediator particle separately.~For spin 1 mediators, we find that two DM-nucleon interactions operators that were not considered in previous studies can actually arise from the non-relativistic reduction of simplified models for spin 1 DM.~We also find that the new operators, denoted here by $\mathcal{O}_{19}$ and $\mathcal{O}_{20}$, can have an important impact on the calculation of DM-nucleus scattering rates, especially for DM masses below 50~GeV.~In this mass range, scattering rates computed by including $\mathcal{O}_{19}$ and $\mathcal{O}_{20}$ or neglecting them differ by up to one order of magnitude in the high energy tail of the nuclear recoil energy spectrum.~Focusing on a simplified version of DARWIN~\cite{Aalbers:2016jon}, we also show the importance of taking into account the new operators $\mathcal{O}_{19}$ and $\mathcal{O}_{20}$ when assessing the compatibility of a direct detection signal with the CMB constraint on the DM relic density.

The article is organised as follows.~In Sec.~\ref{sec:theory} we introduce the simplified model framework studied in this work.~Sec.~\ref{sec:new} focuses on the non-relativistic reduction of the simplified models introduced in Sec.~\ref{sec:theory}, defining the operators $\mathcal{O}_{19}$ and $\mathcal{O}_{20}$ explicitly.~Sec.~\ref{sec:dd} illustrates their impact on the calculation of the expected rate of nuclear recoil events at DM direct detection experiments.~Sec.~\ref{sec:impact} investigates the compatibility between a DM signal at DARWIN and the DM thermal production mechanism in models where $\mathcal{O}_{19}$ and $\mathcal{O}_{20}$ are quantitatively important.~Details on the derivation of the operators $\mathcal{O}_{19}$ and $\mathcal{O}_{20}$ and on the calculation of the associated DM-nucleus scattering cross-sections are provided in the Appendices.

\section{Simplified models for spin 1 DM}
\label{sec:theory}

This section introduces the theoretical framework that we will use to investigate non-relativistic DM-nucleus scattering (Sec.~\ref{sec:dd}) and DM chemical decoupling (Sec.~\ref{sec:impact}) in the case of spin 1 DM.~We will consider two cases separately.~In the first one, a spin 1 DM candidate interacts with quarks through the exchange of a spin 0 mediator particle, as described below
\begin{align}
\label{eq:scalar_lag}
\mathscr{L}_{\phi} &=  \mathscr{L}_{\text{SM}}  - 
\frac{1}{2}\mathcal{X}^\dagger_{\mu\nu}\mathcal{X}^{\mu\nu} + 
m_{X}^2X^\dagger_\mu X^\mu - \frac{\lambda_X}{2}\left(X^\dagger_\mu 
X^\mu\right)^2\nonumber\\
&+\frac{1}{2}\partial_\mu\phi\partial^\mu\phi - \frac{1}{2}m_\phi^2\phi^2 -\frac{m_\phi\mu_1}{3}\phi^3-\frac{\mu_2}{4}\phi^4  \nonumber \\
&-b_1m_X\phi X^\dagger_\mu X^\mu - \frac{b_2}{2}\phi^2X^\dagger_\mu X^\mu \nonumber \\
&- h_1\phi\bar{q}q - ih_2\phi\bar{q}\gamma^5q \,.
\end{align}
The Lagrangian corresponding to this first case, $\mathscr{L}_{\phi}$, extends the Standard Model Lagrangian, $\mathscr{L}_{\text{SM}}$, by kinetic and interaction terms involving the vector field $X_\mu$ describing the spin 1 DM candidate, and the scalar (or pseudoscalar) field $\phi$ associated with the spin 0 mediator.~In Eq.~(\ref{eq:scalar_lag}), $\mathcal{X}_{\mu\nu}$ is the DM field strength tensor, and a dagger denotes Hermitian conjugation.~Without loss of generality, we assume the coupling constants $b_1$, $b_2$, $h_1$, $h_2$, $\mu_1$, $\mu_2$ and $\lambda_X$ to be real, and denote DM and mediator mass by $m_X$ and $m_\phi$, respectively.~In Eq.~(\ref{eq:scalar_lag}), the spinors $q$ describe the Standard Model quarks.~A summation over flavour indexes is understood, and quark interactions are assumed to be flavour diagonal and of universal strength.~The $4\times4$ matrices $\gamma_\mu$, with $\mu=0,\dots,3$, and $\gamma_5$ are the familiar $\gamma$-matrices.

In addition to Eq.~(\ref{eq:scalar_lag}), we also consider the case of a spin 1 DM candidate interacting with quarks through the exchange of a spin 1 mediator particle, as described by the following Lagrangian
\begin{align}
\label{eq:boson_lag}
\mathscr{L}_{G} &= \mathscr{L}_{\text{SM}} 
-\frac{1}{2}\mathcal{X}^{\dagger}_{\mu\nu}\mathcal{X}^{\mu\nu}+m_{X}^2X^{\dagger
}_{\mu}X^{\mu}-\frac{\lambda_{X}}{2}(X_{\mu}^{\dagger}X^{\mu})^2 \nonumber\\
&-\frac{1}{4}\mathcal{G}_{\mu\nu}\mathcal{G}^{\mu\nu}+\frac{1}{2}m_{G}
^2G_\mu^2-\frac{\lambda_G}{4}(G_\mu G^\mu)^2\nonumber \\
&-\frac{b_3}{2}G_\mu^2(X^{\dagger}_\nu X^{\nu}) 
-\frac{b_{4}}{2}(G^{\mu}G^{\nu})(X^{\dagger}_{\mu}X_{\nu}) \nonumber\\ 
&-\left[
ib_{5}X_{\nu}^{\dagger}\partial_{\mu}X^{\nu}G^\mu+b_{6}X_{\mu}^{\dagger
}\partial^\mu X_{\nu}G^{\nu}
+
b_{7}\epsilon_{\mu\nu\rho\sigma}(X^{\dagger\mu}\partial^{\nu}X^{\rho}
)G^ {
\sigma} +\hc \right]\nonumber\\
&-h_3G_\mu\bar{q}\gamma^\mu q - h_4 G_\mu\bar{q}\gamma^\mu\gamma^{5}q\,,
\end{align}
where $m_G$ and $\mathcal{G}_{\mu\nu}$ are, respectively, mass and field strength tensor of the spin 1 mediator $G_\mu$.~Without loss of generality, we assume the coupling constants $\lambda_G$, $b_3$, $b_4$, $b_5$, $h_3$ and $h_4$ in Eq.~(\ref{eq:boson_lag}) to be real, while $b_6$ and $b_7$ are required to be complex, since their phase cannot be reabsorbed by a field redefinition.~As for the case of a spin 0 mediator, we assume flavour diagonal and universal interactions, and a summation over quark flavours is understood.~While the Lagrangians in Eqs.~(\ref{eq:boson_lag}) and (\ref{eq:scalar_lag}) have already been used in the study of DM particle phenomenology at the LHC~\cite{Baum:2017kfa,Baum:2018sxd,Baum:2018lua}, and at direct detection experiments employing ton-scale~\cite{Dent:2015zpa,Catena:2017xqq} or spin-polarised targets~\cite{Catena:2018uae}, in Sec.~\ref{sec:new} we will show that Eqs.~(\ref{eq:boson_lag}) 
predicts new, as yet unexplored DM-nucleon interactions in the non-relativistic limit.~As we will see, the new interactions can have an impact not only on the expected rate of DM-nucleus scattering events at direct detection experiments (Sec.~\ref{sec:dd}), but also on the compatibility between a DM direct detection signal and the CMB constraint on the DM relic density (Sec.~\ref{sec:impact}).

\section{New non-relativistic interactions for spin 1 DM}
\label{sec:new}
The non-relativistic interaction between DM particles and nuclei can be described in terms of nucleon-level quantum mechanical operators~\cite{Fan:2010gt,Fitzpatrick:2012ix}.~In the most general case, these interaction operators can be constructed from the Hermitian and Galilean-invariant objects~\cite{Baum:2017kfa,Baum:2018sxd,Baum:2018lua}:
\begin{equation}
\label{eq:list_of_small_ops}
\mathds{1}_N, \qquad \mathds{1}_X, \qquad \frac{i\vec{q}}{m_N}, \qquad \vt, \qquad \snno, 
\qquad \sXno, \qquad \sy\,,
\end{equation}
where $\mathds{1}_N$ and $\mathds{1}_X$ are identity operators in nucleon and DM spin space, $\vec{q}$ is the momentum transferred in the DM-nucleon interaction, and $\vt \equiv \vec{v} + \vec{q}/(2\mu)$ is the transverse DM-nucleon relative velocity obeying $\vt\cdot\vec{q} = 0$.~Here $\vec{v}$ and $\mu$ are the DM-nucleon relative velocity and reduced mass, respectively.~We denote by $m_N$ the nucleon mass.~The vectors $\snno$ and $\sXno$ are the nucleon and DM spin operators, while the Galilean-invariant $\sy$ is a symmetric combination of DM polarisation vectors.~An explicit expression for $\sy$ is given below in Eq.~(\ref{eq:spin1S}).~Tab.~\ref{tab:operators} shows the operators defining the effective theory of DM-nucleon interactions~\cite{Fan:2010gt,Fitzpatrick:2012ix}.

We obtain the nucleon-level quantum mechanical operators associated with Eqs.~(\ref{eq:scalar_lag}) and (\ref{eq:boson_lag}) as follows.~We first integrate out the particle mediators ($\phi$ in the case of Eq.~(\ref{eq:scalar_lag}) and $G_\nu$ in the case of Eq.~(\ref{eq:boson_lag})), assuming that the momentum transferred in the scattering is much smaller than the mediator mass.~Then we match the resulting quark-level Lagrangians on to relativistic nucleon-level Lagrangians for DM-nucleon interactions by using nucleon form factors from~\cite{Dent:2015zpa}.~Finally, we use the nucleon-level Lagrangians obtained in this manner to compute the corresponding amplitudes for DM-nucleon scattering, which we expand at second order in $|\vec{q}|/m_N$ and first order in $\vec{v}^\perp$.~In the case of a spin 0 mediator, we find
\begin{align}
\label{eq:L0}
i \mathcal{M}^{(0)}&= \frac{2ib_1h_1^Nm_X m_N}{m_\phi^2}\delta^{s's}\delta^{r'r} + \frac{2ib_1h_2^Nm_X m_N}{m_\phi^2}\delta^{s's}\frac{i\vec{q}}{m_N}\cdot\sn \,,
\end{align}
while in the case of a spin 1 mediator described by Eq.~(\ref{eq:boson_lag}), we obtain the non-relativistic amplitude
\begin{align}
\label{eq:L1}
i\mathcal{M}^{(1)}&= -\frac{4ib_5h_3^Nm_X m_N}{m_G^2}\delta^{s's}\delta^{r'r} + \frac{8ib_5h_4^Nm_N m_X}{m_G^2}\delta^{s's}\vt\cdot\sn\nonumber\\
&-\frac{4\Re(b_6)h_3^N}{m_G^2}\bigg\{m_N\vec{q}\cdot\sy^{s's}\cdot\vt\delta^{r'r} - i\left[\vec{q}\cdot\sy^{s's}\cdot(\vec{q}\times\sn)\right]\bigg\} \nonumber\\
&+\frac{8\Re(b_6)h_4^Nm_N}{m_G^2}\left(\vec{q}\cdot\sy^{s's}\cdot\sn\right) \nonumber \\
&-\frac{i\Im(b_6)h_3^N}{m_G^2}\bigg\{-\frac{2m_N}{m_X}\left(\vec{q}\cdot\sy^{s's}\cdot\vec{q}\right)\delta^{r'r} +2im_N\sX\cdot(\vec{q}\times\vt)\delta^{r'r}\nonumber\\
&-2q^2(\sX\cdot\sn) + 2(\vec{q}\cdot\sX)(\vec{q}\cdot\sn)\bigg\} \nonumber\\
&-\frac{4\Im(b_6)h_4^Nm_N}{m_G^2}\sX\cdot(\vec{q}\times\sn)\nonumber\\
 &+\frac{4i\Re(b_7)h_3^Nm_X}{m_G^2}\bigg\{m_N\vt\cdot\sX\delta^{r'r} - i\sX\cdot(\vec{q}\times\sn)\bigg\} \nonumber\\
 &-\frac{8i\Re(b_7)h_4^Nm_Nm_X}{m_G^2}\Big(\sX\cdot\sn\Big) \nonumber\\
 &+\frac{2\Im(b_7)h_3^Nm_N}{m_G^2}\sX\cdot\vec{q}\,\delta^{r'r} \nonumber\\
 &+\frac{i\Im(b_7)h_4^N}{m_G^2}\bigg\{4im_N\left(\vec{q}\cdot\sX\right)\left(\vt\cdot\sn\right) + 4\frac{m_N}{m_X}\vec{q}\cdot\sy^{s's}\cdot(\vec{q}\times\sn)\bigg\} \,,
\end{align}
where $2 \sn=\xi^{\dagger r'} \boldsymbol{\sigma}_N \xi^r$, and $\boldsymbol{\sigma}_N$ is a vector whose components are the three Pauli matrixes acting on the nucleon spinors $\xi^{r}$, $r=1,2$.~Here, we omitted the isospin indexes carried by spinors $\xi^{r}$ and used the following compact notation to simplify the above equations, e.g.,~$h_3^N \delta^{r' r}=\xi^{\dagger r'} h_3^N \xi^r \equiv \xi^{\dagger r'} \left[h_3^p(\mathds{1}+\tau_3)/2 +h_3^n(\mathds{1}-\tau_3)/2\right] \xi^{r} $, where $\mathds{1}$ and $\tau_3$ are the identity and third Pauli matrix acting on the nucleon isospin space.~The couplings to protons, $h_k^p$, and neutrons, $h_k^n$, equal the constants $h_k$, $k=1,\dots,4$, in Eqs.~(\ref{eq:scalar_lag}) and (\ref{eq:boson_lag}) times a nucleon form factor that we take from~\cite{Dent:2015zpa} (and references therein).~In Eq.~(\ref{eq:L1}), we only considered terms at most quadratic in $\vec{q}$ and linear in $\vt$, and expressed the DM particle spin $\sX$ and the $\sy^{s' s}$ operator in terms of three-dimensional DM polarisation vectors $e_{si}$ and $e'_{s'j}$, 
\begin{align}
\label{eq:spin1S}
\sX &= -i\esp\times\es\,,\nonumber\\
\sy^{s's}_{ij} &= \frac{1}{2}\left(e_{si}e'_{s'j} + e_{sj}e'_{s'i}\right)\,.
\end{align}
\begin{table}[t]
    \centering
    \begin{tabular*}{\columnwidth}{@{\extracolsep{\fill}}ll@{}}
    \toprule
        $\mathcal{O}_1 = \mathds{1}_{X}\mathds{1}_N$  & $\mathcal{O}_{11} = i\vec{S}_X\cdot\frac{\vec{q}}{m_N}\mathds{1}_N$    \\
        $\mathcal{O}_3 = i\vec{S}_N\cdot\left(\frac{\vec{q}}{m_N}\times\vt\right)\mathds{1}_X$ & $\mathcal{O}_{12} = \vec{S}_X\cdot \left(\vec{S}_N \times\vt \right)$  \\
        $\mathcal{O}_4 = \vec{S}_X\cdot \vec{S}_N$ & $\mathcal{O}_{13} =i \left(\vec{S}_X\cdot \vt\right)\left(\vec{S}_N\cdot \frac{\vec{q}}{m_N}\right)$ \\
        $\mathcal{O}_5 = i\vec{S}_X\cdot\left(\frac{\vec{q}}{m_N}\times\vt\right)\mathds{1}_N$ &  $\mathcal{O}_{14} = i\left(\vec{S}_X\cdot \frac{\vec{q}}{m_N}\right)\left(\vec{S}_N\cdot \vt\right)$ \\
        $\mathcal{O}_6 = \left(\vec{S}_X\cdot\frac{\vec{q}}{m_N}\right) \left(\vec{S}_N\cdot\frac{\vec{q}}{m_N}\right)$ &$\mathcal{O}_{15} = -\left(\vec{S}_X\cdot \frac{\vec{q}}{m_N}\right)\left[ \left(\vec{S}_N\times \vt \right) \cdot \frac{\vec{q}}{m_N}\right] $  \\
        $\mathcal{O}_7 = \vec{S}_N\cdot \vt\mathds{1}_X$ & $\mathcal{O}_{17}=i \frac{\vec{q}}{m_N} \cdot \boldsymbol{\mathcal{S}} \cdot \vt \mathds{1}_N$  \\
        $\mathcal{O}_8 = \vec{S}_X\cdot \vt\mathds{1}_N$ & $\mathcal{O}_{18}=i \frac{\vec{q}}{m_N} \cdot \boldsymbol{\mathcal{S}}  \cdot \vec{S}_N$  \\
        $\mathcal{O}_9 = i\vec{S}_X\cdot\left(\vec{S}_N\times\frac{\vec{q}}{m_N}\right)$ & $\mathcal{O}_{19} = \frac{\vec{q}}{m_N} \cdot \boldsymbol{\mathcal{S}} \cdot \frac{\vec{q}}{m_N}$  \\
       $\mathcal{O}_{10} = i\vec{S}_N\cdot\frac{\vec{q}}{m_N}\mathds{1}_X$   & $\mathcal{O}_{20} = \left(\vec{S}_N \times \frac{\vec{q}}{m_N} \right) \cdot \boldsymbol{\mathcal{S}} \cdot \frac{\vec{q}}{m_N}$  \\
    \bottomrule
    \end{tabular*}
    \caption{List of quantum mechanical operators defining the non-relativistic effective theory of DM-nucleon interactions~\cite{Fan:2010gt,Fitzpatrick:2012ix,Dent:2015zpa}, augmented by including the operators $\mathcal{O}_{19}$ and $\mathcal{O}_{20}$ that we found to arise from the non relativistic reduction of simplified models for spin 1 DM with a vector mediator (see Eq.~(\ref{eq:boson_lag})).}
\label{tab:operators}
\end{table}
\begin{table}[t]
    \centering
    \begin{tabular*}{\columnwidth}{@{\extracolsep{\fill}}ll@{}}
    \toprule
$c_1^N = 
-\dfrac{b_5h_3^N}{m_G^2}$ & \\ 
$c_4^N = \dfrac{\Im(b_6)h_3^Nq^2}{2 m_G^2m_Xm_N} - \dfrac{2\Re(b_7)h_4^N}{m_G^2}$  & $c_{11}^N = -\dfrac{\Im(b_7)h_3^N m_N}{2m_G^2m_X}$ \\
$c_5^N = -\dfrac{\Im(b_6)h_3^N m_N}{2m_G^2m_X}$  & $c_{14}^N = \dfrac{\Im(b_7)h_4^N m_N}{m_G^2m_X}$ \\
$c_6^N = -\dfrac{\Im(b_6)h_3^N m_N}{2m_G^2m_X}$ & $c_{17}^N = \dfrac{\Re(b_6)h_3^N m_N}{m_G^2m_X}$ \\
$c_7^N = \dfrac{2b_5h_4^N}{m_G^2}$  & $c_{18}^N = -\dfrac{2\Re(b_6)h_4^N m_N}{m_G^2m_X}$ \\
$c_8^N = \dfrac{\Re(b_7)h_3^N}{m_G^2}$  & $c_{19}^N = \dfrac{\Im(b_6)h_3^N m_N^2}{2m_G^2 m_X^2}$ \\
$c_9^N = \dfrac{\Re(b_7)h_3^N}{m_G^2} - \dfrac{\Im(b_6)h_4^N m_N}{m_G^2m_X}$   & $c_{20}^N = -\dfrac{\Re(b_6)h_3^N m_N}{m_G^2m_X} - \dfrac{\Im(b_7)h_4^Nm_N^2}{m_G^2m_X^2}$ \\ 
\bottomrule
\end{tabular*}
\caption{Non zero coupling constants appearing in Eq.~(\ref{eq:LNR}).~They arise from the non-relativistic reduction of Eq.~(\ref{eq:boson_lag}).~The coupling constants $h_k^N$, $k=1,2,3,4$, with $N=n$ for neutrons and $N=p$ for protons, equal the coupling constants $h_k$ times a nucleon form factor that we take from~\cite{Dent:2015zpa} (and references therein).~Notice that the coupling constants in this table are a factor of 2 smaller than the ones in~\cite{Catena:2017xqq,Baum:2017kfa}.~There, we build $\mathscr{M}_{{\rm NR}}$ (defined in Appendix~\ref{sec:app_derivation}) following the prescription outlined in~\cite{Dent:2015zpa}.}
\label{tab:spin1_coeffs_list}
\end{table}

\noindent In a reference frame where the $z$-axis is in the direction of the three-dimensional DM particle momentum, the four-dimensional DM polarisation vectors, $\epsilon^{s\mu}(p)$ and $\epsilon^{s'\mu*}(p')$, can be written as
\begin{align}
\label{eq:def_pol_states}
\epsilon^{s\mu}(p) = \begin{pmatrix}\dfrac{|\vec{p}\,|}{m_X}\delta^{3s}\\[5mm]
\dfrac{p_0}{m_X}\vec{e}_{s}\end{pmatrix}\,; \qquad\epsilon^{s'\mu*}(p') = \begin{pmatrix}\dfrac{|\vec{p}\,'|}{m_X}\delta^{3s'}\\[5mm]
\dfrac{p'_0}{m_X}\vec{e}\,'_{s'}\end{pmatrix}\,,
\end{align}
where $s=1,2,3$, $\vec{e}_3=\vec{p}/|\vec{p}|$ and $\vec{e}\,'_{3}=\vec{p}\,'/|\vec{p}\,'|$.~From the amplitudes in Eqs.~(\ref{eq:L0}) and (\ref{eq:L1}), one can read the nucleon-level quantum mechanical operators associated with Eqs.~(\ref{eq:scalar_lag}) and (\ref{eq:boson_lag}), respectively.~For example, the amplitude in Eq.~(\ref{eq:L1}) can be rewritten in a more compact form using the operators in Tab.~\ref{tab:operators} and the nucleon-level coupling constants in Tab.~\ref{tab:spin1_coeffs_list},
\begin{align}
\label{eq:LNR}
\mathcal{M}^{(1)} = 4 m_X m_N \sum_{i} \left[ c^{p}_i \, \left\langle\mathcal{O}_i \, \frac{(\mathds{1}+\tau_3)}{2} \right\rangle+ c^{n}_i \,  \left\langle\mathcal{O}_i \, \frac{(\mathds{1}-\tau_3)}{2} \right\rangle \right]\,,
\end{align}
where the sum runs over the non zero coefficients in Tab.~\ref{tab:spin1_coeffs_list}, and angle brackets denote expectation values between initial and final DM and nucleon spin and isospin states.~The amplitude in Eq.~(\ref{eq:L0}) can also be written as in Eq.~(\ref{eq:LNR}), but in this case only two operators contribute to the sum and the corresponding coupling constants are given by:~$c_1^N =b_1h_1^N/(2m_\phi^2)$ and $c_{10}^N = b_1h_2^N/(2m_\phi^2)$.

The operators $\mathcal{O}_{19}$ and $\mathcal{O}_{20}$ in Eq.~(\ref{eq:LNR}) were not included in previous classifications, and their phenomenology will be investigated here for the first time.~Explicitly, they read as follows
\begin{align}
\mathcal{O}_{19} &= \frac{\vec{q}}{m_N} \cdot \boldsymbol{\mathcal{S}} \cdot \frac{\vec{q}}{m_N}\,, \nonumber\\
\mathcal{O}_{20} &= \left(\vec{S}_N\ \times \frac{\vec{q}}{m_N} \right) \cdot \boldsymbol{\mathcal{S}} \cdot \frac{\vec{q}}{m_N}\,.
\end{align}
They are quadratic in the momentum transfer, linear in the symmetric combination of polarisation vectors for the DM particle, $\boldsymbol{\mathcal{S}}$, and specific to spin 1 DM.~As we will see in the next sections, they are not negligible in the analysis of DM direct detection experiments, since they arise from simplified models which do not generate momentum transfer independent operators as leading interactions in the non-relativistic expansion of DM-nucleon scattering amplitudes.

\section{New interactions and non-relativistic DM-nucleus scattering}
\label{sec:dd}

\subsection{Expected rate of DM-nucleus scattering events}

The rate per unit detector mass of DM-nucleus scattering events in a direct detection experiment is
\begin{equation}
\label{eq:rate}
\frac{{\rm d}R}{{\rm d}E_{\text{R}}} = \frac{\rho_\chi}{m_X m_A}\int {\rm d}^3v\, vf(\vec{v}) \frac{{\rm d}\sigma}{{\rm d}E_{\text{R}}}\,,
\end{equation}
where $f(\vec{v})$ is the DM velocity distribution in the detector rest frame, $\rho_\chi$ is the local DM density, $m_A$ is the mass of the target nuclei, $E_R$ is the nuclear recoil energy, and ${\rm d}\sigma/{\rm d}E_{\text{R}}$ is the differential cross-section for DM-nucleus scattering.~For the local DM density and velocity distribution in the detector rest frame we assume, respectively, $\rho_\chi=0.4$~GeV~cm$^{-3}$ and
\begin{align}
\label{eq:velodis}
f(\vec{v}) &=  (\pi v_0^2)^{-1}\left[\sqrt{\pi}v_0\erf\left(\frac{v_{\text{esc}}}{v_0}\right) - 2v_{\text{esc}}e^{-v_{\text{esc}}^2/v_0^2}\right]^{-1}
e^{-(\vec{v} + \vec{v}_{\text{e}})^2/v_0^2}\,; \qquad |\vec{v} + \vec{v}_{\text{e}}|\le v_{\rm esc}  \,,
\end{align}
where $\vec{v}_{\text{e}}$ ($v_{\text{esc}}=544$~km~s$^{-1}$) is the Earth (escape) velocity in the galactic rest frame,  and the most probable speed, $v_0=220$~km~s$^{-1}$, equals the circular speed of the local standard of rest.~The differential cross-section for DM-nucleus scattering in Eq.~(\ref{eq:rate}) can be expressed as follows
\begin{equation}
\label{eq:diff_cross_section1}
\frac{{\rm d}\sigma}{{\rm d}E_R} = \frac{m_A}{2 \pi v^2} |\overline{T_{fi}}|^2,
\end{equation}
where $T_{fi}$ is the matrix element between initial and final DM-nucleus states of the underlying nuclear potential, and a bar denotes average (sum) over initial (final) DM and nuclear spin configurations.~We calculate the matrix element $T_{fi}$ for the cases of a spin 0 and a spin 1 mediator separately.~To this end, we first calculate the inverse Fourier transforms of $\mathcal{M}^{(0)}/(4 m_X m_N)$ and $\mathcal{M}^{(1)}/(4 m_X m_N)$ and sum the result over all constituent nucleons to obtain the position-space nuclear potential associated with the DM and mediator model under investigation.~Then we evaluate the matrix element between initial and final DM-nucleus states of the obtained nuclear potential to obtain $T_{fi}$.~Following~\cite{Anand:2013yka}, for $T_{fi}$ we find
\begin{equation}
\label{eq:matrix_element2}
    T_{fi}= \sum_{\tau = 0,1}\bra{f}
    \ell_{0}^{\tau} 
S^{\tau} + \ell_{0}^{A \tau} T^{\tau} + \vec{\ell}_{5}^{\tau}\cdot \vec{P}^{\tau} +  
\vec{\ell}_{M}^{\tau}\cdot \vec{Q}^{\tau} +  \vec{\ell}_{E}^{\tau}\cdot 
\vec{R}^{\tau}
\ket{i}
\end{equation}
where initial and final nuclear states are denoted by $\ket{i}$ and $\ket{f}$, respectively.~Nuclear charge operators ($S^{\tau}$ and $T^{\tau}$) and nuclear vector operators ($\vec{P}^{\tau}$, $\vec{Q}^{\tau}$, and $\vec{R}^{\tau}$) are defined in Appendix~\ref{sec:nuc}.~They arise from the expansion in spherical harmonics of the matrix element between initial and final DM states of the position-space nuclear potential obtained from $\mathcal{M}^{(0)}$ and $\mathcal{M}^{(1)}$.~Summation over the constituent nucleons is encoded in the definition of the nuclear charges and currents.~The sum over $\tau$ extends from 0 (isoscalar-coupling) to 1 (isovector coupling).~Isoscalar and isovector contributions to Eq.~(\ref{eq:matrix_element2}) are proportional to the identity and the third Pauli matrix in isospin space, respectively.~For further details, see Appendix~\ref{sec:nuc}.~From Eqs.~(\ref{eq:scalar_lag}) and (\ref{eq:boson_lag}), for the $\ell$-coefficients multiplying $S^{\tau}$, $T^{\tau}$, $\vec{P}^{\tau}$, $\vec{Q}^{\tau}$, and $\vec{R}^{\tau}$ we find
\begin{align}
\label{eq:ell_coeff}
\ell_0^\tau &=
c_{1}^\tau\delta^{s's}
+ i\ct{5} \left(\qmn\times\vt_T\right)\cdot\sX
+ \ct{8}\left(\vt_T\cdot\sX\right) 
+ i\ct{11}\left(\qmn\cdot\sX\right)  \nonumber
\\
& + i\ct{17}\left(\qmn\cdot\sy^{s's}
\cdot\vt_T\right) + \ct{19}\left(\qmn\cdot\sy^{s's}
\cdot\qmn\right) 
\,, \nonumber\\
\ell_0^{A\tau} &= 
-\frac{1}{2}\ct{7}\delta^{s's}
- i\ct{14}\left(\qtmn\cdot\sX\right)
\,, \nonumber\\
\vec{\ell}_E^\tau &= 0 
\,, \nonumber\\
\vec{\ell}_M^\tau &= 
i\ct{5}\left(\qmn\times\sX\right)
-\ct{8} \sX
-i\ct{17}\left(\qmn\cdot\sy^{s's}
\right)
\,, \nonumber \\
\vec{\ell}_5^\tau &= 
\frac{1}{2}\ct{4}\sX + 
\frac{1}{2}\ct{6}\left(\qmn\cdot\sX\right)\qmn
+ \frac{1}{2}\ct{7}\vt_T\delta^{s's}
+ \frac{i}{2}\ct{9}\left(\qmn\times\sX\right)
+ \frac{i}{2}\ct{10}\qmn\delta^{s's}  
\nonumber\\
&
+\frac{i}{2}\ct{14}\left(\qmn\cdot\sX\right)\vt_T
+\frac{i}{2}\ct{18}\left(\qmn\cdot\sy^{s's}
\right)
-\frac{1}{2}\ct{20}\left[\left(\qmn\cdot\sy^{s's}
\right)\times\qmn\right]\,, \nonumber\\
\end{align}
where $s$ and $s'$ label, respectively, the initial and final DM particle spin state, $\vt_T=\vec{v}+\vec{q}/(2\mu_T)$, and $\mu_T$ is the reduced DM-nucleus mass.~Here $\vec{v}$ is the relative velocity between the incoming DM particle and a reference point in the target nucleus, e.g.~its centre of mass.~The velocity of individual nucleons with respect to this reference point explicitly appears in the definition of nuclear charges and currents given in Appendix~\ref{sec:nuc}~\cite{Fitzpatrick:2012ix}.~With this notation, the squared modulus of $T_{fi}$ admits the following decomposition in terms of nuclear response functions, $W^{\tau \tau'}_k(q^2)$, $k=M, \Sigma', \Sigma'', \Phi'', \tilde{\Phi}', \Delta M, \Phi '', \Sigma' \Delta$, DM response functions, $R^{\tau \tau'}_k(q^2,v^2)$, $k=M, \Sigma', \Sigma'', \Phi'', \tilde{\Phi}', \Delta M, \Phi '', \Sigma' \Delta$ (defined in~\cite{Fitzpatrick:2012ix}), and target nucleus spin $J$~\cite{Anand:2013yka}:
\begin{equation}
\label{eq:matrixRW}
    |\overline{T_{fi}}|^2= \frac{4\pi}{2J+1}
\sum_k\sum_{\tau = 0,1}\sum_{\tau' = 0,1}\left(\frac{q}{m_N}\right)^{2\ell(k)}R_k^{\tau\tau'}W_k^{\tau\tau'}.
\end{equation}
where $\ell(k)=0$ for $k=M, \Sigma', \Sigma''$ and $\ell(k)=1$ otherwise.~In our calculations we use nuclear response functions from \texttt{DMFormFactor}~\cite{Anand:2013yka} for the most abundant xenon isotopes.~Their normalisation is such that $4\pi/(2J+1)W^{00}_M(0)=A^2/4$, where $A$ is the atomic mass number.~This normalisation implies the relations: $c_j^p=(c_j^0+c_j^1)/2$ and $c_j^n=(c_j^0-c_j^1)/2$, where the coupling constants for protons, $c_j^p$, and neutrons, $c_j^n$ are listed in Tab.~\ref{tab:spin1_coeffs_list}.~The normalisation $4\pi/(2J+1)W^{00}_M(0)=A^2$ would have implied $c_j^p=(c_j^0+c_j^1)$ and $c_j^n=(c_j^0-c_j^1)$.~Using Eq.~(\ref{eq:ell_coeff}) and Eq.~(\ref{eq:Rl2}), for the DM response functions in Eq.~(\ref{eq:matrixRW}) we find the following expressions
\begin{align}
\label{eq:DMresponses}
R^{\tau\tau'}_{M} &= \ct{1}\ctp{1} + \frac{2}{3}\vt_T\,^2\qpmn{2}\ct{5}\ctp{5} + 
\frac{2}{3}\vt_T\,^2\ct{8}\ctp{8} + \frac{2}{3}\qpmn{2}\ct{11}\ctp{11} + 
\frac{1}{6}\qpmn{2}\vt_T\,^2\ct{17}\ctp{17}\nonumber\\
&
+ \frac{1}{3}\qpmn{4}\ct{19}\ctp{19} +\frac{1}{3}\qpmn{2}\left(\ct{1}\ctp{19} + 
\ct{19}\ctp{1} \right) \nonumber\\
R^{\tau\tau'}_{\Sigma''} &=  \frac{1}{6}\ct{4}\ctp{4} + 
\frac{1}{6}\qpmn{2}\left(\ct{4}\ctp{6} + \ct{6}\ctp{4} \right) + 
\frac{1}{6}\qpmn{4}\ct{6}\ctp{6} + \frac{1}{4}\qpmn{2}\ct{10}\ctp{10} \nonumber\\
&+ \frac{1}{12}\qpmn{2}\left(\ct{10}\ctp{18} + \ct{18}\ctp{10} \right)
+ \frac{1}{12}\qpmn{2}\ct{18}\ctp{18} \nonumber\\
R^{\tau\tau'}_{\Sigma'} &=  \frac{1}{6}\ct{4}\ctp{4} + 
\frac{1}{8}\vt_T\,^2\ct{7}\ctp{7} + \frac{1}{6}\qpmn{2}\ct{9}\ctp{9} + 
\frac{1}{12}\qpmn{2}\vt_T\,^2\ct{14}\ctp{14} + 
\frac{1}{24}\qpmn{2}\ct{18}\ctp{18}  \nonumber\\
& + \frac{1}{24}\qpmn{4}\ct{20}\ctp{20}  \nonumber\\
R^{\tau\tau'}_{\Delta} &= \frac{2}{3}\qpmn{2}\ct{5}\ctp{5} + 
\frac{2}{3}\ct{8}\ctp{8}+ \frac{1}{6}\qpmn{2}\ct{17}\ctp{17} \nonumber\\
R^{\tau\tau'}_{\Delta\Sigma'} &= \frac{2}{3}\ct{5}\ctp{4} - 
\frac{2}{3}\ct{8}\ctp{9} + \frac{1}{6}\qpmn{2}\ct{17}\ctp{20} \nonumber\\
R^{\tau\tau'}_{\Phi''} &=  0 \,, \qquad 
R^{\tau\tau'}_{\Phi''M} = 0 \,, \qquad 
R^{\tau\tau'}_{\tilde{\Phi}'} =  0 \,.
\end{align}
These are the most general response functions arising from Eqs.~(\ref{eq:scalar_lag}) and (\ref{eq:boson_lag}) in the case of spin 1 DM.~While in the case of Eq.~(\ref{eq:boson_lag}) all coupling constants for DM-nucleon interactions in Eqs.~(\ref{eq:ell_coeff}) and (\ref{eq:DMresponses}) can be different from zero (they are listed in Tab.~\ref{tab:spin1_coeffs_list}), in the case of Eq.~(\ref{eq:scalar_lag}), only the coupling constants $c_1^\tau$ and $c_{10}^\tau$ are different from zero (as one can see from Eq.~(\ref{eq:L0})).~Notice also that the interference term proportional to $(\ct{10}\ctp{18}+\ct{18}\ctp{10})$ can only be different from zero when DM couples to quark via scalar and vector mediator simultaneously.

\begin{figure}[t]
    \centering
    \includegraphics[width=.4\linewidth]{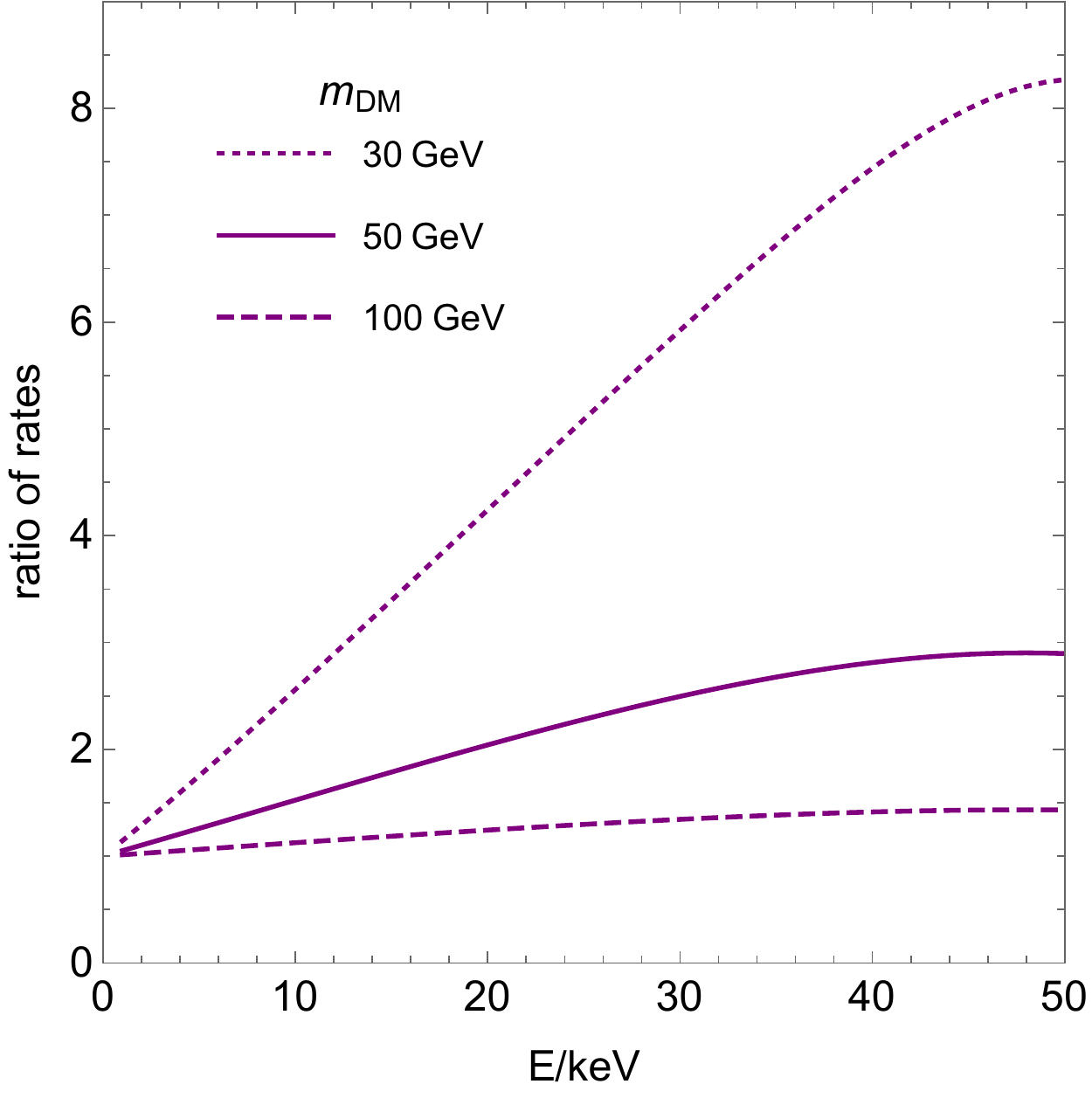}
    \includegraphics[width=.423\linewidth]{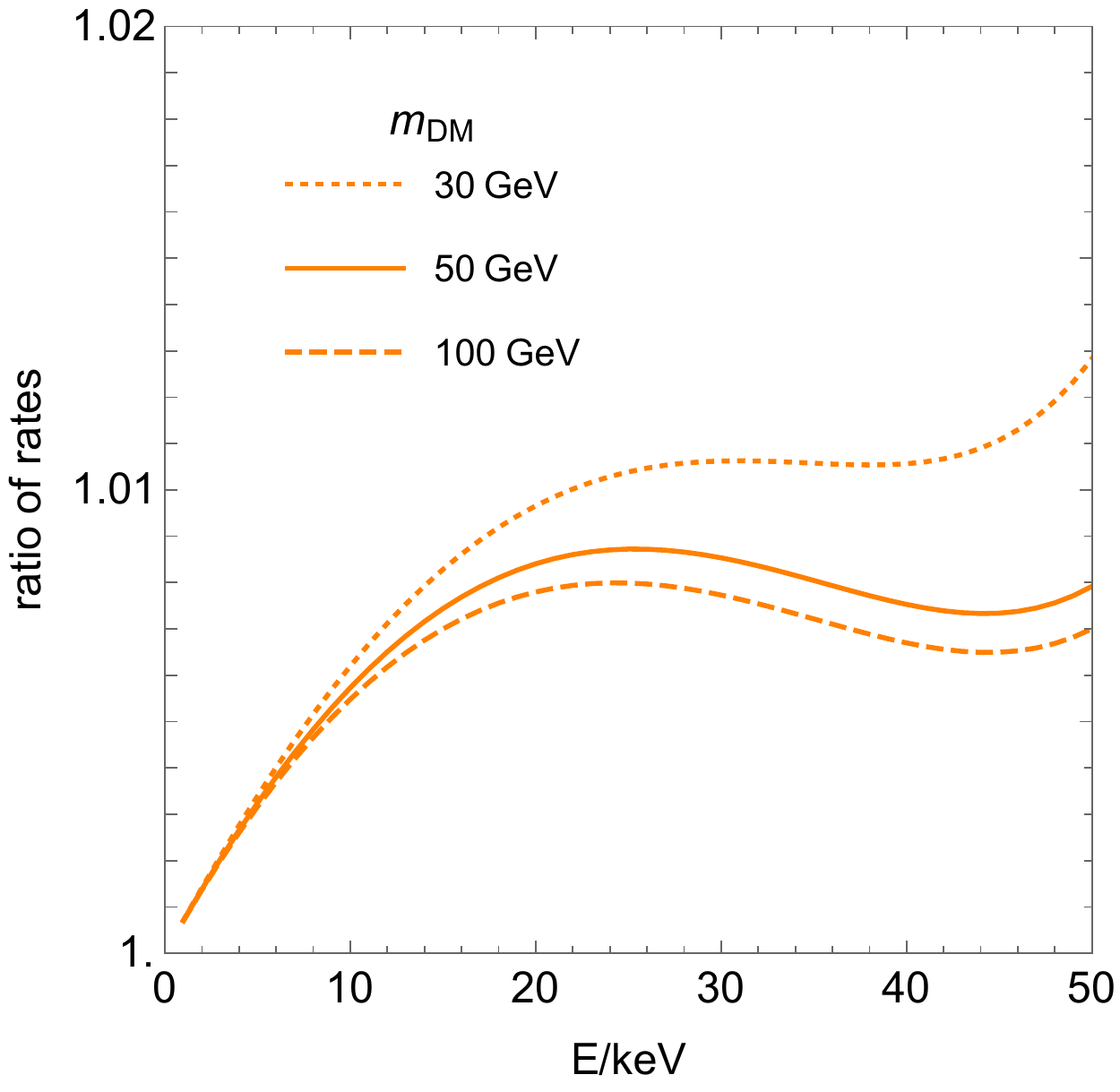}
     \includegraphics[width=.4\linewidth]{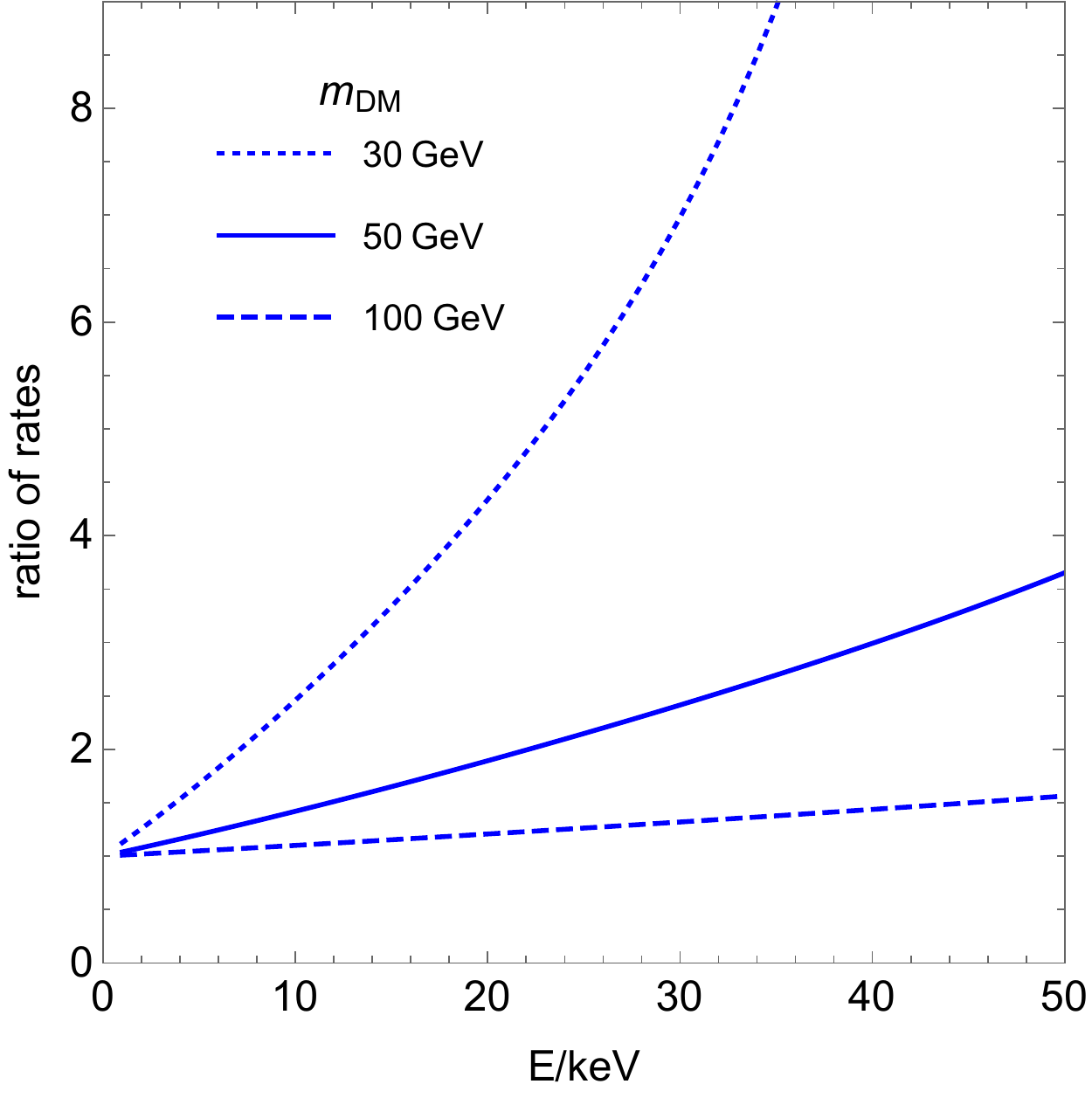}
      \includegraphics[width=.4205\linewidth]{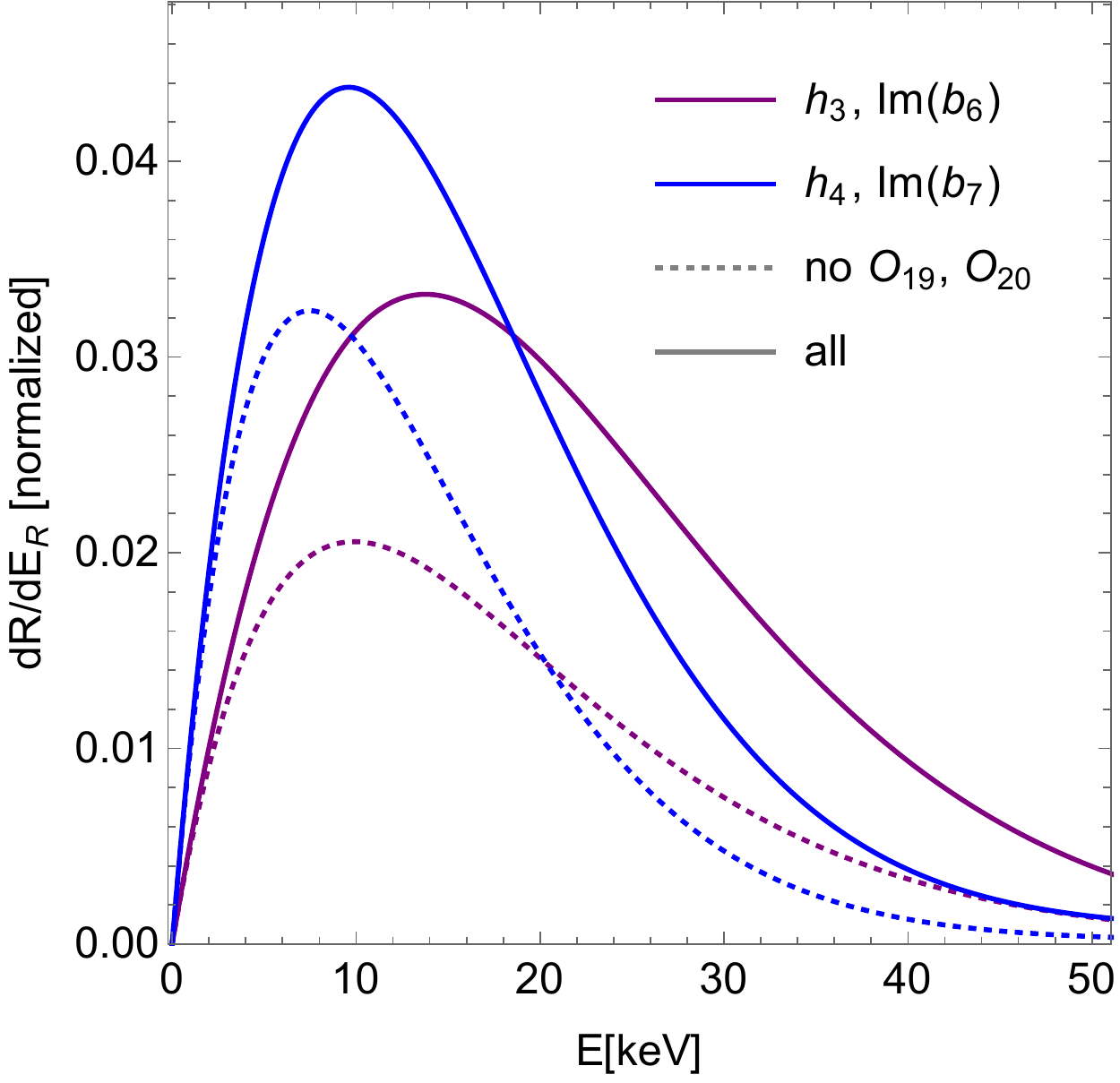}
    \caption{Expected rate of DM-nucleus scattering events for the models ($h_3$, $\Im(b_6)$), top left panel, ($h_3$, $\Re(b_6)$), top right panel, and ($h_4$, $\Im(b_7)$), bottom left panel.~In all panels, the exact rate (computed by taking into account contributions from the new operators $\mathcal{O}_{19}$ or $\mathcal{O}_{20}$) is divided by an approximate scattering rate computed by erroneously neglecting $\mathcal{O}_{19}$ or $\mathcal{O}_{20}$.~The bottom right panel of Fig~\ref{fig:rates} shows the absolute rate of expected nuclear recoils for the models ($h_3$, $\Im(b_6)$) and ($h_4$, $\Im(b_7)$).~Solid (dashed) lines correspond to the exact (approximate) calculation.~Interestingly, nuclear recoil energy spectra computed neglecting the operator $\mathcal{O}_{19}$ or $\mathcal{O}_{20}$ peak at smaller nuclear recoil energies.}
    \label{fig:rates}
\end{figure}
\subsection{Numerical results}
\label{sec:num1}
In this section we numerically evaluate the expected rate of DM-nucleus scattering events, Eq.~(\ref{eq:rate}), assuming xenon as a target material, and focusing on three benchmark models separately:~1) In a first model, $h_3$ and $\Im(b_6)$ are the only coupling constants different from zero in Eq.~(\ref{eq:boson_lag});~2) In a second model $h_3$ and $\Re(b_6)$ are different from zero while all other couplings are zero;~3) In the third model all couplings in Eq.~(\ref{eq:boson_lag}) are equal to zero, but $h_4$ and $\Im(b_7)$.~In the three models, DM and the particle mediator have spin 1.~When $h_3$ is different from zero, DM couples to the vector quark current, whereas when $h_4$ is different from zero DM couples to the axial quark current.~Depending on the model, the DM-mediator vertex is determined by the $b_6$ or $b_7$ coupling.~The first model generates $\mathcal{O}_{19}$, the other ones $\mathcal{O}_{20}$.

For the three simplified models described above, Fig.~\ref{fig:rates} shows the expected rate of DM-nucleus scattering events (computed by taking into account contributions from the new operators $\mathcal{O}_{19}$ or $\mathcal{O}_{20}$) divided by an ``approximate'' scattering rate, computed by erroneously neglecting $\mathcal{O}_{19}$ or $\mathcal{O}_{20}$.~In this figure, the top left panel refers to the ($h_3$, $\Im(b_6)$) model, the top right panel to model ($h_3$, $\Re(b_6)$) and, finally, the bottom left panel to the model ($h_4$, $\Im(b_7)$).

Fig.~\ref{fig:rates} shows that $\mathcal{O}_{19}$ cannot be neglected in analyses of model ($h_3$, $\Im(b_6)$) and, similarly, that $\mathcal{O}_{20}$ cannot be neglected when computing scattering rates for model ($h_4$, $\Im(b_7)$).~In both cases, we find that neglecting the operator $\mathcal{O}_{19}$ or $\mathcal{O}_{20}$ leads to significantly underestimate the expected rate of DM-nucleus scattering events, especially for $m_X<50$~GeV and $E_R$ larger than about $20$~keV.~At the same time, in the case of model ($h_3$, $\Re(b_6)$) the ratio of rates in the top right panel of Fig.~\ref{fig:rates} is close to one at all recoil energies and for all masses.

The bottom right panel of Fig~\ref{fig:rates} shows the absolute rate of expected nuclear recoils for the models ($h_3$, $\Im(b_6)$) and ($h_4$, $\Im(b_7)$).~For each model, we calculate the absolute rate exactly (including contributions from $\mathcal{O}_{19}$ or $\mathcal{O}_{20}$) and ``approximately'' (neglecting contributions from $\mathcal{O}_{19}$ or $\mathcal{O}_{20}$).~Solid lines correspond to the exact calculation, whereas dashed lines represent the ``approximated'' results.~Interestingly, for both models we find that nuclear recoil energy spectra computed neglecting $\mathcal{O}_{19}$ or $\mathcal{O}_{20}$ peak at smaller nuclear recoil energies.~Therefore, not only the total number of expected signal events is larger when $\mathcal{O}_{19}$ and $\mathcal{O}_{20}$ are properly taken into account, but also the spectral distribution of the events is appreciably modified.

\section{New interactions and DM chemical decoupling}
\label{sec:impact}

\subsection{Expected DM relic density}
\noindent In this section, we briefly review how to compute the DM relic density starting from Eqs.~(\ref{eq:scalar_lag}) and (\ref{eq:boson_lag}).~We refer to~\cite{Catena:2017xqq,Gondolo:1990dk} (and references therein) for further details on this calculation.~Assuming that DM was in kinetic and chemical equilibrium in the early Universe, the time evolution of its cosmological number density, $n$, is described by the Boltzmann equation,
\begin{equation}
\label{eq:boltzmann1}
    \dot{n} + 3Hn = -\langle\sigma_{\rm ann} v_{\text{M\o l}}\rangle\left(n^2 - 
n^2_{\text{eq}}\right),
\end{equation}
where $H$ is the Hubble rate in the assumed cosmological model, $\sigma_{\rm ann}$ is the invariant DM pair annihilation cross section, $n_{eq}$ is the temperature dependent DM equilibrium number density, and
\begin{equation}
    v_{\text{M\o l}} = \frac{\sqrt{(p_1\cdot p_2)^2 - m_X^4}}{E_1E_2}
\end{equation}
is the M\o ller velocity.~In Eq.~(\ref{eq:boltzmann1}), angle brackets denote an average over the DM thermal distribution, which can be expressed in terms of the lab frame DM-DM relative velocity, $v_{\text{lab}}$, as follows~\cite{Gondolo:1990dk}
\begin{equation}
\label{eq:simgavlab}
    \langle\sigma_{\rm ann} v_{\text{M\o l}}\rangle = \int_0^{\infty}\text{d}\epsilon 
\frac{2x}{K_2^2(x)}\epsilon^{1/2}(1 + 2\epsilon)K_1(2x\sqrt{1 + x})\sigma_{\rm ann} 
v_{\text{lab}}\,.
\end{equation}
In the above expression, $\epsilon = (s - 4m_X^2)/4m_X^2$ is the total kinetic energy per unit mass in the lab frame, $s = (p_1 + p_2)^2$ is the squared centre-of-mass energy for the annihilation process, $x = m_X/T$, where $T$ is the CMB temperature, and $K_1(x)$ and $K_2(x)$ are the first two modified Bessel functions of the second kind.~The product $\sigma_{\rm ann} v_{\rm lab}$ can be computed as follows 
\begin{equation}
    \sigma_{\rm ann} v_{\text{lab}} = \frac{1}{64\pi^2(s - 2m_X^2)}\left[1 - \frac{(m_3 + 
m_4)^2}{s}\right]^{1/2}\left[1 - \frac{(m_3 - m_4)^2}{s}\right]^{1/2}\int 
\text{d}\Omega \overline{|\mathcal{M}|}^2,
\end{equation}
where $m_3$ and $m_4$ are the masses of the annihilation products, ${\rm d}\Omega$ is an infinitesimal solid angle in the center-of-mass frame, and $\mathcal{M}$ is the invariant amplitude for DM pair annihilation.~We calculate the invariant amplitude $\mathcal{M}$ within the theoretical framework introduced in Sec.~\ref{sec:theory}, starting from Eqs.~(\ref{eq:scalar_lag}) and (\ref{eq:boson_lag}).~The Boltzmann equation in Eq.~(\ref{eq:boltzmann1}) can be rewritten in terms of the DM abundance $Y = n/\mathcal{S}$, where $\mathcal{S}$ is the total entropy density of the Universe,
\begin{equation}
\label{eq:boltzmann2}
    \frac{\text{d}Y}{\text{d}x} = 
-\left(\frac{45}{\pi}G\right)^{-1/2}\frac{g_*^{1/2}m_X}{x^2}\langle\sigma 
v_{\text{M\o l}}\rangle\left(Y^2 - Y^2_{\text{eq}}\right).
\end{equation}
Here, $g_*^{1/2}$ is the degrees of freedom parameter defined as
\begin{equation}
\label{eq:gstar}
    g_*^{1/2} = \frac{h_{\text{eff}}}{g_{\text{eff}}^{1/2}}\left(1 + 
\frac{1}{3}\frac{T}{h_{\text{eff}}}\frac{\text{d}h_{\text{eff}}}{\text{d}T}
\right),
\end{equation}
and the equilibrium density is given by
\begin{equation}
\label{eq:yeq}
    Y_{\text{eq}} = \frac{45gx^2K_2(x)}{4\pi^4h_{\text{eff}}(m_X/x)},
\end{equation}
where $g$ is the number of internal degrees of freedom for the DM particle. The quantities $h_{\text{eff}}$ and $g_{\text{eff}}$ in Eq.~(\ref{eq:gstar}) and 
(\ref{eq:yeq}) are effective degrees of freedom for the entropy and energy densities, respectively~\cite{Gondolo:1990dk}.~An approximate solution to the DM abundance equation, Eq.~(\ref{eq:boltzmann2}), is given by
\begin{equation}
    \frac{1}{Y_0} = \frac{1}{Y_{\text{f}}} + 
\left(\frac{45}{\pi}G\right)^{1/2}\int_{T_0}^{T_{\text{f}}}g_*^{1/2}
\langle\sigma v_{\text{M\o l}}\rangle\text{d}T \,,
\end{equation}
where $Y_0$ is the present value of the DM abundance (the so-called DM relic abundance), $Y_{\rm f}$ is the DM abundance at the freeze-out temperature, $T_{\rm f}$, and $T_{\rm f}$ is defined as the temperature below which $Y\neq Y_{\rm eq}$.~The freeze-out temperature can be found approximately by solving for $T$~\cite{Gondolo:1990dk} 
\begin{equation}
\left(\frac{45}{\pi}G\right)^{1/2}\frac{45g}{4\pi^4}\frac{K_2(x)}{h_{\text{eff}}
(T)}g_*^{1/2}m_X\langle\sigma v_{\text{M\o l}}\rangle\delta(\delta + 2) = 
\frac{K_1(x)}{K_2(x)}\,,
\end{equation}
where $\delta\equiv(Y - Y_{\text{eq}})/Y_{\text{eq}}$ is set to 1.5~\cite{Gondolo:1990dk}.~Finally, the DM relic density in critical units reads as follows
\begin{equation}
    \Omega_{\text{DM}}h^2 = 2.8282\times 
10^8\left(\frac{m_X}{\text{GeV}}\right)\left(\frac{T_0}{2.75\text{K}}
\right)^3Y_0,
\end{equation}
where $h = H_0/100$, $H_0$ is the Hubble constant, and $T_0$ is the present value of the CMB temperature.

\begin{figure}[t]
    \centering
    \includegraphics[width=.488\linewidth]{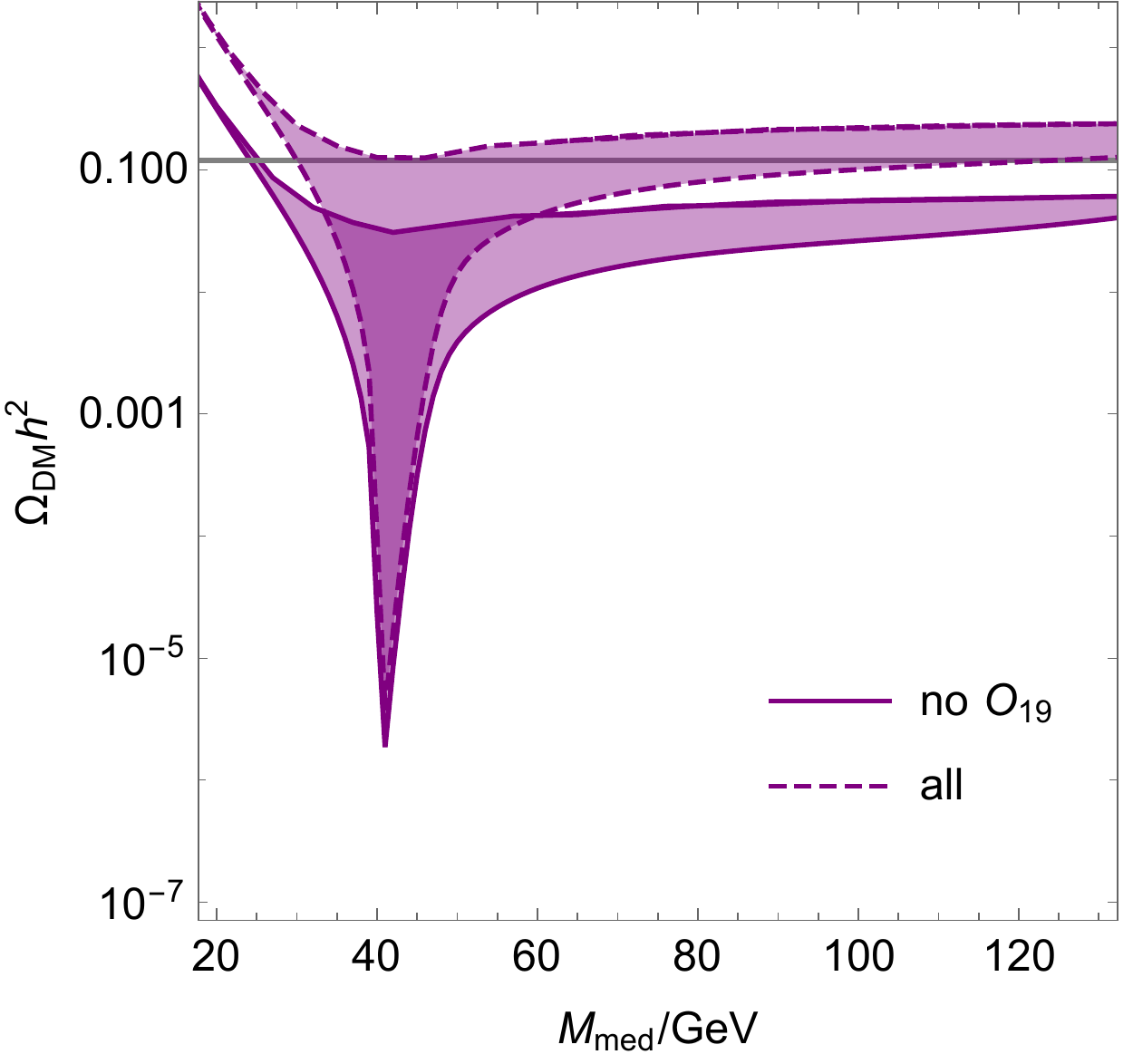}
    \includegraphics[width=.48\linewidth]{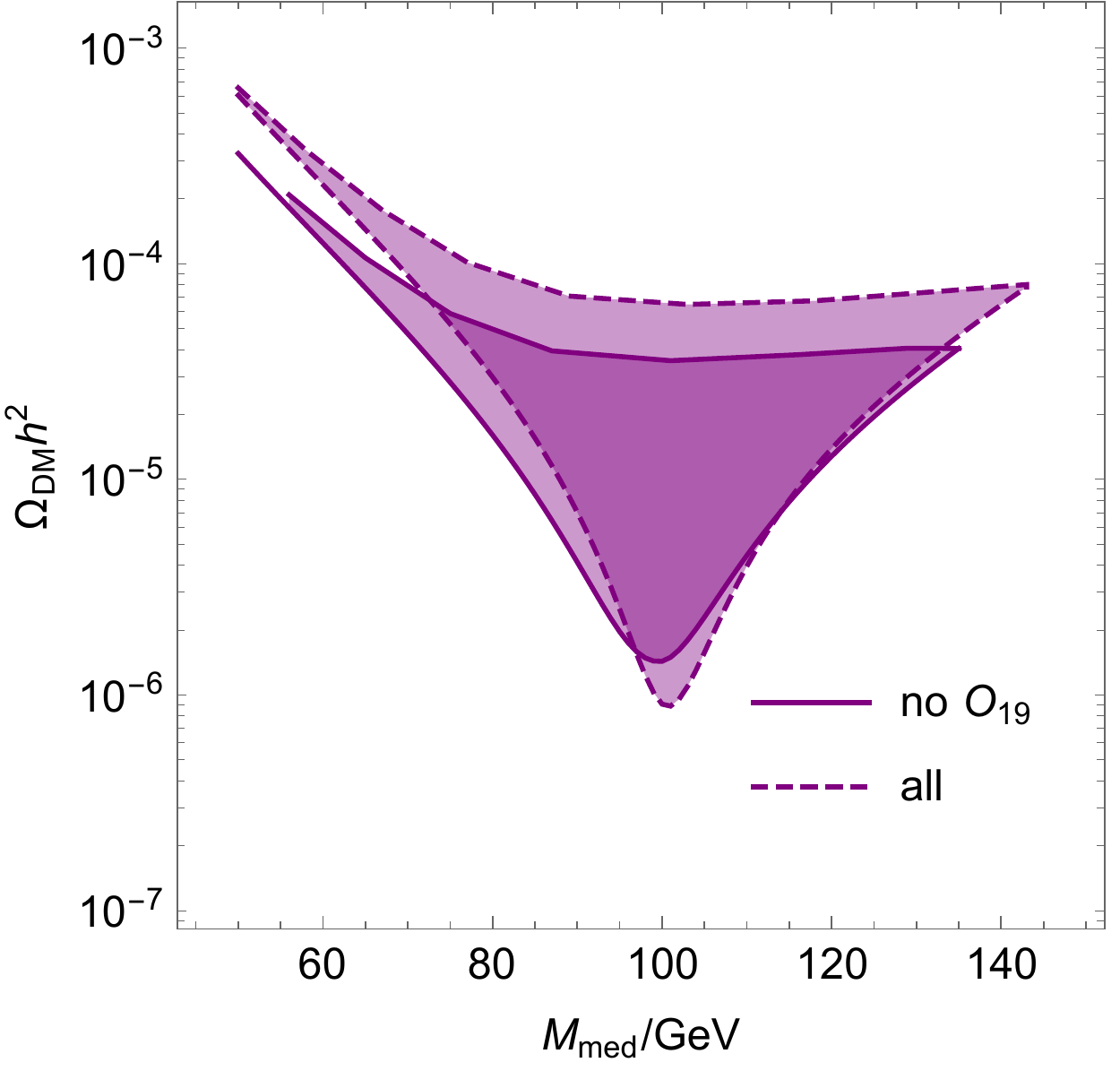}
    \caption{Relic density calculation for model ($h_3$, $\Im(b_6)$).~Regions in the ($m_{\rm med}$, $\Omega_{\rm DM} h^2$) plane that are compatible with $N_{\rm obs}=1$ and $m_X^*=20$~GeV (left panel) and $N_{\rm obs}=150$ and $m_X*=50$~GeV (right panel) when assuming an exposure of 80 ton$\times$year and the interval [5 keV, 45 keV] as a signal region for the DARWIN experiment.~Here $N_{\rm obs}$ is the number of observed signal events, and $m_X^*$ a benchmark value for the DM mass.~We compare exact results taking into account the $\mathcal{O}_{19}$ contribution to the observed signal (dashed contours) with approximate results which neglect the latter (solid contours).}
    \label{fig:relic}
\end{figure}

\subsection{Numerical results}
In this section we investigate the compatibility of a signal in a future direct detection experiment, such as DARWIN, with current CMB constraints on the DM relic density in scenarios where the new operators $\mathcal{O}_{19}$ or $\mathcal{O}_{20}$ are important.~We start by calculating the DM relic density for the simplified model characterised by $h_3$ and $\Im(b_6)$ as non zero coupling constants.~We perform this calculation setting the model parameters to values producing $N=N_{\rm obs}$ signal events at DARWIN, and keeping the DM particle mass fixed to the benchmark value $m_X=m_X^*$ (which we assume to be reconstructed from the DARWIN signal).~The constraints $N=N_{\rm obs}$ and $m_X=m_X^*$ (together with the assumption that coupling constants are less than $\sqrt{4 \pi}$, as required by perturbativity) identify a two-dimensional surface in the four-dimensional space spanned by the model parameters $m_X$, mediator mass $m_{\rm med}=m_G$, $h_3$ and $\Im(b_6)$.~If the projection of this two-dimensional surface onto the ($m_{\rm med}$, $\Omega_{\rm DM} h^2$) plane intersects the relic density constraint, $\Omega_{\rm DM} h^2=0.12$~\cite{Ade:2015xua}, then the ($h_3$, $\Im(b_6)$) model is compatible with the observation of $N=N_{\rm obs}$ signal events at DARWIN.

Fig.~\ref{fig:relic} shows the regions in the ($m_{\rm med}$, $\Omega_{\rm DM} h^2$) plane that are compatible with $N_{\rm obs}=1$ and $m_X^*=20$~GeV (left panel) and $N_{\rm obs}=150$ and $m_X^*=50$~GeV (right panel) when assuming the interval [5 keV, 45 keV] as a signal region and an exposure of 80 ton$\times$year for DARWIN.~In both panels, we compare exact results that take into account the $\mathcal{O}_{19}$ contribution to the observed DARWIN signal (dashed contours) with ``approximate'' results which neglect the latter (solid contours).~We find that neglecting the contribution from $\mathcal{O}_{19}$ to the expected number of signal events in DARWIN when assessing the compatibility of DARWIN signal and CMB constraints can lead to wrong conclusions, especially for small values of $N_{\rm obs}$.~For example, in the case of $N_{\rm obs}=1$ (or, analogously, in a scenario where DARWIN has recorded the first ``few'' signal events) the range of mediator masses within which DARWIN signal and CMB constraints are compatible varies from a narrow window at small mediator masses (neglecting $\mathcal{O}_{19}$) to a much broader interval (taking $\mathcal{O}_{19}$ into account).

In this section we presented the results of our numerical calculations focusing on the ($h_3$, $\Im(b_6)$) model.~However, we find simular results for model ($h_4$, $\Im(b_7)$), which in the non-relativistic limit generates the $\mathcal{O}_{20}$ operator.~At the same time, in the case of the ($h_3$, $\Re(b_6)$) model, which also generates the $\mathcal{O}_{20}$ operator in the non-relativistic limit, we find that $\mathcal{O}_{20}$ can safely be ignored when assessing the compatibility of a DARWIN signal with CMB constraints.

\section{Beyond simplified models for spin 1 DM}
\label{sec:discussion}
Inspection of the amplitudes in Eqs.~(\ref{eq:L0}) and (\ref{eq:L1}) shows that the only operators depending on the Galilean invariant $\sy$ that can arise from the non-relativistic reduction of simplified models for spin 1 DM are $\Op_{17}$ and $\Op_{18}$, and the new operators $\Op_{19}$ and $\Op_{20}$.~However, from an effective theory perspective it is interesting to ask what is the general structure of the Hermitian and Galilean invariant DM-nucleon interaction operators that can be build by means of $\sy$.

At second order in the momentum transfer $\vec{q}$ and at first order in the transverse relative velocity $\vt$ and in $\sy$, we find that four additional non-relativistic operators (compared to the ones already listed in Tab.~\ref{tab:operators}) can be constructed from the Galilean invariants in Eq.~(\ref{eq:list_of_small_ops}), namely
\begin{align}
       \Op_{21} &= \vt\cdot\sy\cdot\vec{S}_N \,, &
       \Op_{22} &= \left(i\qmn \times\vt\right)\cdot\sy\cdot\vec{S}_N \,,
    \nonumber\\
       \Op_{23} &= i\qmn\cdot\sy\cdot\left(\vec{S}_N\times\vt\right) \,,& 
       \Op_{24} &=  \vt\cdot\sy\cdot\left(\vec{S}_N \times i\qmn \right) \,.
\label{eq:four}
\end{align}
These operators do not appear in the non-relativistic reduction of known 
simplified models for DM interactions, but 
they are Galilean invariant and Hermitian, and are therefore not forbidden a priori. 

\section{Conclusion}
\label{sec:conclusion}
In this work, we focused on the non-relativistic reduction of simplified models for spin 1 DM.~We explored two cases separately:~in the first case, DM couples to quarks via the exchange of a spin 0 mediator; in the second one, the interactions of DM with quarks are mediated by a spin 1 particle.~Our goal was to identify features in the phenomenology of simplified models for DM-quark interactions which are specific to spin 1 DM.~We were especially interested in simplified model predictions for DM direct detection and the DM relic density.~To calculate the expected rate of nuclear recoils in DM direct detection experiments, for each simplified model we expanded the amplitude for DM-nucleon scattering at second order in the momentum transfer (and at first order in the transverse relative velocity), in analogy with previous works on spin 0 and spin 1/2 DM~\cite{Fan:2010gt,Fitzpatrick:2012ix}.~We then calculated the inverse Fourier transform of the non-relativistic amplitude, summed over all constituent nucleons, and obtained the position-space nuclear potential associated with the underlying DM and mediator model.~We finally evaluated the matrix element of the obtained potential between initial and final DM-nucleus states to find the non-relativistic cross section for DM-nucleus scattering.~To obtain the DM relic density from a given simplified model, we calculated the cross section for DM pair annihilation into quarks using the formalism in~\cite{Gondolo:1990dk}.

In the case of spin 1 mediators, we found two DM-nucleon interaction operators arising from the non-relativistic reduction of simplified models for spin 1 DM that were not considered in previous studies and are specific to spin 1 DM.~They are quadratic in the momentum transfer and depend on a symmetric combination of the DM particle polarisation vectors.~Exploring the phenomenology of the new operators, denoted here by $\mathcal{O}_{19}$ and $\mathcal{O}_{20}$, we found that these can have an important impact on the expected rate of DM-nucleus scattering events at DM direct detection experiments, especially when DM is lighter than 50~GeV, and despite being quadratic in the momentum transfer.~This is related to the fact that they arise from simplified models which do not generate momentum transfer independent operators at leading order in the non-relativistic expansion.~For example, we found that nuclear recoil energy spectra computed by including $\mathcal{O}_{19}$ and $\mathcal{O}_{20}$ or neglecting them differ by up to one order of magnitude for nuclear recoil energies larger than about 20 keV when $m_X<50$~GeV.~Furthermore, we found that the new operators can induce appreciable distortions in the expected nuclear recoil energy spectra, such as a shift in the corresponding peaks. 

Focusing on a simplified version of the DARWIN experiment, we investigated under what circumstances a signal at DARWIN can be compatible with constraints from CMB data on the DM relic density in simplified models where $\mathcal{O}_{19}$ and $\mathcal{O}_{20}$ are important.~We found that, when assessing the compatibility of a signal at DARWIN with CMB constraints, neglecting the contribution from $\mathcal{O}_{19}$ to the expected number of signal events in DARWIN can lead to inaccurate conclusions, especially when the number of signal events at DARWIN, $N_{\rm obs}$, is small (e.g.~$\mathcal{O}(1)$).~For example, for $N_{\rm obs}=1$ the range of mediator masses within which DARWIN signal and CMB constraints are compatible varies from a narrow region at small mediator masses, when $\mathcal{O}_{19}$ is neglected, to a broad region, when $\mathcal{O}_{19}$ is taken into account.

Finally, we concluded our analysis with a brief discussion on the non-relativistic operators for DM-nucleon interactions which can be built in terms of the symmetric combination of DM polarisation vectors $\sy$, but which do not arise from the non-relativistic reduction of simplified models for spin 1 DM.~We found that, at second order in the momentum transfer and at first order in the transverse relative velocity and in $\sy$, four additional non-relativistic operators can be built from the Galilean invariants in Eq.~(\ref{eq:list_of_small_ops}).~While these additional operators do not appear in the non-relativistic limit of the simplified models introduced in Sec.~\ref{sec:theory}, they are Galilean invariant and Hermitian (see Eq.~(\ref{eq:four})) and therefore not forbidden a priori. 

\section*{Acknowledgments}
This work is performed within the Swedish Consortium for Dark Matter Direct Detection (SweDCube), and was supported by the Knut and Alice Wallenberg Foundation (PI, Jan Conrad) and by the Vetenskapsr\aa det (Swedish Research Council), contract No. 638-2013-8993 (PI, Katherine Freese).

\appendix

\section{Derivation of the new interaction operators: $\mathcal{O}_{19}$ and $\mathcal{O}_{20}$}
\label{sec:app_derivation}

In this appendix we explicitly derive the operators $\mathcal{O}_{19}$ and $\mathcal{O}_{20}$ from the non-relativistic reduction of Eq.~(\ref{eq:boson_lag}).~The first step in this derivation is to ``integrate out'' the heavy mediator $G_\mu$ to obtain an effective Lagrangian that is valid at energies below the mediator mass, $m_G$.~This is done by replacing the vector field $G_\mu$ in Eq.~(\ref{eq:boson_lag}) with the solution of the equation of motion 
\begin{equation}
 \left[ 
 \frac{\partial \mathcal{L}}{\partial G_\mu} 
 - \partial_\nu \frac{\partial \mathcal{L}}{\partial \partial_\nu G_\mu}
 \right]_{G_\mu = G_\mu^0} =0\,,
\end{equation}
where $G_\mu=G_\mu^0$, is the leading order solution in a $|\vec{q}|/m_G$ expansion.~Assuming DM-mediator and quark-mediator interactions of the general form $\mathscr{L}_{\rm int}=G^\mu A_\mu$, where $A_\mu$ can be any of the DM and quark bilinears appearing in Eq.~(\ref{eq:boson_lag}), a simple calculation gives the following result
\begin{equation}
 \mathscr{L_\text{eff}} = - \frac{1}{2 m_G^2} A^\mu A_\mu\,.
 \label{eq:Leff}
\end{equation}
Here, we will separately focus on the three $A^\mu$ bilinear combinations that can contribute to the new operators $\mathcal{O}_{19}$ and $\mathcal{O}_{20}$.

\subsection{Mediator interaction via $\Re(b_6)$ and $h_3$}
In this first scenario, the contact DM-nucleon interactions are described by Eq.~(\ref{eq:Leff}) and the $A^\mu$ bilinear
\begin{align}
\label{eq:A1}
 A^\mu &= - \Re(b_6)(X^{\dagger\nu} \partial_\nu X^\mu + \hc) - h_3^N \bar N \gamma^\mu N \,,
 \end{align}
where the spinor field $N$ describes the nucleon.~From Eqs.~(\ref{eq:Leff}) and (\ref{eq:A1}), we calculate the amplitude for DM-nucleon scattering, $\mathcal{M}^{(1)}=4 m_X m_N \mathscr{M}_{\rm NR}$, which in the non-relativistic limit is given by 
\begin{align}
\label{eq:M1}
\mathscr{M}_{\rm NR} =
-\frac{\Re(b_6)h_3^N}{2 m_G^2 m_X} i q^\nu
 (\epsilon_\nu^{s'*} \epsilon^s_\mu + \epsilon_\mu^{s'*} \epsilon^s_\nu)
 \left\{
 \delta^{\mu0}\delta^{r'r} + g^{\mu m} \frac{1}{2m_N} 
 [\vec{K}\delta^{r' r} - 2i(\sn \times \vec{q})]_m
 \right\} \,,
\end{align}
where 
\begin{align}
\label{eq:epsilon}
  \epsilon^{s \mu}(\vec{p}) &\simeq
  \begin{pmatrix}
   \frac{1}{2 m_X}(\vec{P}-\vec{q}) \cdot \vec{e}_s \\
   \vec{e}_s
  \end{pmatrix} \,, \nonumber \\
   \epsilon^{s'\mu *}(\vec{p}')  &\simeq
  \begin{pmatrix}
   \frac{1}{2 m_X}(\vec{P}+\vec{q}) \cdot \vec{e}_{s'}' \\
   \vec{e}_{s'}'
  \end{pmatrix}\,.
\end{align}
Here $\vec{P} = \vec{p} + \vec{p'} = 2 \vec{p} + \vec{q}$ is the sum of 
incoming and outgoing DM momenta and the vectors $\epsilon^{s \mu}$ ($\vec{e}_s$) 
are the four (three) dimensional polarization vectors of the DM field.~We also used the non-relativistic reduction of the free nucleon spinor, $u_N$, which at leading order in $\vec{q}$ gives 
\begin{equation}
 \bar{u}_N^{r'} \gamma^\mu u_N^{r}
 \simeq
 \begin{pmatrix}
  2 m_N \delta^{r' r}\\
  \vec{K}\delta^{r' r} - 2i(\sn \times \vec{q})
 \end{pmatrix}\,,
\end{equation}
where $\vec{K} = \vec{k} + \vec{k'} = 2\vec{k} - \vec{q}$ is the the 
sum of the incoming and outgoing momenta of the 
nucleon.~Finally $q^\nu=(E_{\vec{p}'}-E_{\vec{p}},\vec{p}'-\vec{p})$, where $E_{\vec{p}}$ and $E_{\vec{p}'}$ are initial and final DM particle energy, respectively.~Expanding the contracted indices in Eq.~(\ref{eq:A1}), and using Eq.~(\ref{eq:epsilon}) and the identity 
\begin{align}
-2 m_X q^\nu (\epsilon_\nu^{s'*} \epsilon^{s}_\mu + \epsilon_\mu^{s'*} 
\epsilon^{s}_\nu) \delta^{\mu0} 
&= 
(\vec{q}\cdot\vec{e}'_{s'})[(\vec{P}-\vec{q})\cdot\vec{e}_s] + 
[(\vec{P}+\vec{q})\cdot\vec{e}'_{s'}](\vec{q}\cdot\vec{e}_{s}) 
\nonumber \\
& \equiv 2 \vec{q}\cdot \sy^{s' s} \cdot \vec{P}\,,
\end{align}
at leading order in $\vec{q}$ we find
\begin{align}
  \mathscr{M}_\text{NR}=
 &
\frac{\Re(b_6)h_3^N}{2 m_G^2 m_X} 
\left[
\frac{i}{m_X}\vec{q}\cdot \sy^{s' s} \cdot \vec{P}\delta^{r'r}
- \frac{i}{m_N} \vec{q} \cdot \sy^{s' s} \cdot \vec{K}\delta^{r'r}
- \frac{2 \vec{q}}{m_N} \cdot \sy^{s' s} \cdot (\sn \times \vec{q})
\right]\,.
\end{align}
Finally, by using the relation $\vec{v}^\perp =\vec{P}/(2m_X) - \vec{K}/(2m_N)$ we obtain the non-relativistic amplitude
\begin{align}
 \mathscr{M}_\text{NR} =
\frac{\Re(b_6)h_3^N}{m_G^2 m_X} 
\left[
i\vec{q}\cdot \sy^{s's} \cdot \vec{v}^\perp\delta^{r'r}
- \frac{\vec{q}}{m_N} \cdot \sy^{s's}
\cdot (\sn \times \vec{q})
\right]\,.
\label{eq:Leff1}
\end{align}
Comparing Eq.~(\ref{eq:Leff1}) with Tab.~\ref{tab:operators}, we find that the above amplitude can also be written as follows
\begin{align}
  \mathscr{M}_\text{NR} = \frac{\Re(b_6)h_3^N m_N}{m_G^2 m_X} 
\left( \left\langle\mathcal{O}_{17}\right\rangle - \left\langle\mathcal{O}_{20}\right\rangle \right)\,,
\end{align}
where angle brackets denote expectation values between initial and final DM and nucleon spin and isospin states.

\subsection{Mediator interaction via $\Im(b_6)$ and $h_3$}

In this second scenario, the contact DM-nucleon interactions are described by Eq.~(\ref{eq:Leff}) and the $A^\mu$ bilinear
\begin{align}
\label{eq:Aaux}
 A^\mu = - \Im(b_6)(i X^{\dagger\nu} \partial_\nu X^\mu + \hc)- h_3^N \bar N \gamma^\mu N\,.
 \end{align}
By replacing the bilinear $A^\mu$ in Eq.~(\ref{eq:Aaux}) in to the effective interaction Lagrangian in Eq.~(\ref{eq:Leff}), for the non-relativistic amplitude for DM-nucleon scattering we find the following expression
\begin{equation}
 \mathscr{M}_\text{NR} =
\frac{\Im(b_6)h_3^N}{2 m_G^2 m_X} q^\nu
 (\epsilon_\nu^{s'*} \epsilon^s_\mu - \epsilon_\mu^{s'*} \epsilon^s_\nu)
 \left\{
 \delta^{\mu0}\delta^{r'r} + g^{\mu m} \frac{1}{2m_N} 
 [\vec{K}\delta^{r'r} - 2i(\sn \times \vec{q})]_m
 \right\} \,.
\end{equation}
By expanding the contracted indices in the above expression, at leading order in $\vec{q}$ we obtain the amplitude
\begin{align}
\label{eq:aux}
 \mathscr{M}_\text{\rm NR} &=
-\frac{\Im(b_6)h_3^N}{2 m_G^2 m_X} 
\left\{
\frac{1}{2m_X}(\vec{q}\cdot \vec{e}'_{s'})[(\vec{P}-\vec{q})\cdot\vec{e}_s] \delta^{r'r}
-\frac{1}{2 m_X}(\vec{q}\cdot \vec{e}_s)[(\vec{P}+\vec{q})\cdot\vec{e}'_{s'}]\delta^{r'r}
\right. \nonumber\\ & \left.
- \frac{1}{2m_N} (\vec{q}\cdot \vec{e}'_{s'})(\vec{K}\cdot\vec{e}_s)\delta^{r'r}
+\frac{1}{2m_N}(\vec{q}\cdot \vec{e}_s)(\vec{K}\cdot\vec{e}'_{s'})\delta^{r'r}
\right. \nonumber\\ & \left.
+ \frac{i}{m_N} (\vec{q}\cdot \vec{e}'_{s'})[(\sn\times\vec{q})\cdot\vec{e}_s]
-\frac{i}{m_N}(\vec{q}\cdot \vec{e}_s)[(\sn\times\vec{q})\cdot\vec{e}'_{s'}]
\right\}\,.
\end{align}
Using 
$(\vec{a}\times\vec{b})\cdot(\vec{c} \times\vec{d}) = 
(\vec{a}\cdot\vec{c})(\vec{b}\cdot\vec{d})-(\vec{a}\cdot\vec{d} )(\vec{b} 
\cdot\vec{c})$, where $\vec{a}, \vec{b}, \vec{c}$ and $\vec{d}$ are three-dimensional vectors, Eq.~(\ref{eq:aux}) reads
\begin{align}
 \mathscr{M}_\text{\rm NR} &=
-\frac{\Im(b_6)h_3^N}{ 2 m_G^2 m_X} 
\left\{
\frac{i}{2m_X}(\vec{q}\times\vec{P})\cdot \sX\delta^{r'r}
-\frac{1}{m_X} \vec{q} \cdot \sy^{s's} \cdot \vec{q} \delta^{r'r}
- \frac{i}{2m_N} (\vec{q} \times \vec{K}) \cdot \sX\delta^{r'r}
\right. \nonumber \\ & \left.
-\frac{1}{m_N} \left[
|\vec{q}|^2 (\sX \cdot \sn) - 
(\vec{q}\cdot\sX)(\vec{q}\cdot\sn)
\right]
\right\}\,.
\end{align}
After expressing $\vec{P}$ in terms of $\vt$ and $\vec{K}$, we find that terms proportional to $\vec{K}$ cancel, and obtain
\begin{align}
\label{eq:Leff2}
\mathscr{M}_\text{\rm NR}&=
-\frac{\Im(b_6)h_3^N}{ 2 m_G^2 m_X} 
\left\{
i(\vec{q}\times\vec{v}^\perp)\cdot \sX \delta^{r'r}
-\frac{1}{m_X} \vec{q} \cdot \sy^{s's} \cdot \vec{q} \,\delta^{r'r}
\right. \nonumber \\ & \left.
-\frac{1}{m_N} \left[
|\vec{q}|^2 (\sX \cdot \sn) - 
(\vec{q}\cdot\sX)(\vec{q}\cdot\sn)
\right]
\right\} \,.
\end{align}
Finally, by comparing Eq.~(\ref{eq:Leff1}) with Tab.~\ref{tab:operators}, we find that the above amplitude can also be written as 
\begin{align}
\mathscr{M}_\text{\rm NR}=
\frac{\Im(b_6)h_3^N}{ 2 m_G^2 m_X} \left(
\frac{|\vec{q}^2|}{m_N}\left\langle\mathcal{O}_4\right\rangle
- m_N \left\langle\mathcal{O}_5\right\rangle
- m_N\left\langle\mathcal{O}_6\right\rangle
+ \frac{m_N^2}{m_X} \left\langle\mathcal{O}_{19}\right\rangle
\right) \,.
\end{align}
The term proportional to $\mathcal{O}_{19}$ was missing in previous derivations, and can be numerically important.

\subsection{Mediator interaction via $\Im(b_7)$ and $h_4$}

In this last scenario, the contact DM-nucleon interactions are described by Eq.~(\ref{eq:Leff}) and the $A_\mu$ bilinear 
\begin{align}
 A_\mu &= \Im(b_7)(i\epsilon_{\mu\nu\rho\sigma}X^{\dagger\nu} \partial^\rho X^\sigma + \hc) - h_4^N \bar N \gamma_\mu \gamma_5 N \,.
\end{align}
The non-relativistic amplitude for DM-nucleon scattering, $\mathscr{M}_{\rm NR}$ (defined as above), is thus given by
\begin{equation}
\label{eq:aux2}
 \mathscr{M}_\text{NR} =
\frac{\Im(b_7)h_4^N}{2 m_G^2 m_X} \epsilon_{\mu\nu\rho\sigma} q^\nu 
\epsilon^{s'\rho *} \epsilon^{s \sigma} 
 \left[
 \delta^{\mu0} \sn \cdot  \frac{\vec{K}}{m_N} 
 + 2 g^{\mu m} (\sn)_m
 \right] \,.
\end{equation}
Deriving the above expression, we used the non-relativistic expansion of the axial nucleon spinor bilinear
\begin{equation}
\bar{u}^{r'}_N\gamma^\mu \gamma^5 u^r_N
 \simeq
 \begin{pmatrix}
2   \sn \cdot  \vec{K}   \vspace{0.15 cm} \\ 
  4 m_N\sn
 \end{pmatrix}\,.
\end{equation}
Expanding the contracted indices in Eq.~(\ref{eq:aux2}), we obtain the following non-relativistic amplitude
\begin{align}
\label{eq:aux3}
  \mathscr{M}_\text{NR} &=
\frac{\Im(b_7)h_4^N}{2 m_G^2 m_X} 
\left\{
-\frac{i}{m_N}(\vec{q} \cdot \sX) (\sn \cdot \vec{K})
+ 2 i q^0 (\sX \cdot \sn)
\right . \nonumber\\ &\left.
+ \frac{1}{m_X} (\vec{q} \times \sn) 
\cdot \left[
\vec{e}_s\left(\vec{e}'_{s'}\cdot(\vec{P} + \vec{q})\right) - 
\vec{e}'_{s'}\left(\vec{e}_s\cdot(\vec{P} - \vec{q})\right)
\right]
\right\} \,.
\end{align}
Using the next-to-leading order relation $q^0 = \vec{P} \cdot \vec{q}/(2 m_X)$ and standard vector identities, Eq.~(\ref{eq:aux3}) reads
\begin{align}
 \mathscr{M}_\text{NR} &=
\frac{\Im(b_7)h_4^N}{2 m_G^2 m_X} 
\left\{
-\frac{i}{m_N}(\vec{q} \cdot \sX) (\sn \cdot \vec{K}) 
+ \frac{i}{m_X} (\vec{P} \cdot \vec{q}) (\sX \cdot \sn)
\right . \nonumber\\ & \left.
+ \frac{i}{m_X} (\vec{q} \times \sn) \cdot (\sX \times \vec{P})
+ \frac{2}{m_X} (\vec{q} \times \sn) \cdot \sy^{s's} \cdot \vec{q} 
\right\} \,,
\end{align}
which can be further simplified by combining second and third term.~This leads to the amplitude
\begin{align}
 \mathscr{M}_\text{NR} 
& =
\frac{\Im(b_7)h_4^N}{2 m_G^2 m_X} 
\left\{
-\frac{i}{m_N}(\vec{q} \cdot \sX) (\sn \cdot \vec{K}) 
+ \frac{i}{m_X} (\vec{q} \cdot \sX) (\sn \cdot \vec{P}) \right. \nonumber\\
&\left.  + \frac{2}{m_X} (\vec{q} \times \sn) \cdot \sy^{s's}  \cdot \vec{q} 
\right\}\,.
\end{align}
Finally, the terms proportional to $\vec{K}$ and $\vec{P}$ can be rewritten in terms of a single term proportional to $\vt$,
\begin{align}
\label{eq:Leff3}
  \mathscr{M}_\text{NR} &=
\frac{\Im(b_7)h_4^N}{2 m_G^2 m_X} 
\left\{
2i(\vec{q} \cdot \sX) (\sn \cdot \vec{v}^\perp) 
+ \frac{2}{m_X} (\vec{q} \times \sn) \cdot \sy^{s's}  \cdot \vec{q} 
\right\} \,.
\end{align}
Comparison between Eq.~(\ref{eq:Leff3}) and Tab.~\ref{tab:operators} shows that we can rewrite the above Lagrangian as follows
\begin{align}
  \mathscr{M}_\text{NR} &=
\frac{\Im(b_7)h_4^N}{2 m_G^2 m_X} 
\left(
2 m_N \left\langle\mathcal{O}_{14}\right\rangle
-\frac{2m_N^2}{m_X}\left\langle\mathcal{O}_{20}\right\rangle
\right)\,.
\end{align}
Collecting the results found in this section, for the new operators $\mathcal{O}_{19}$ and $\mathcal{O}_{20}$ we obtain the following coefficients
\begin{align}
c_{19} &= \frac{\Im(b_6)h_3 m_N^2}{2 m_G^2 m_X^2}\,, \nonumber\\
c_{20} &= -\frac{\Re(b_6)h_3 m_N}{m_G^2m_X} 
-\frac{\Im(b_7)h_4m_N^2}{m_G^2m_X^2} \,.
\end{align}

\section{From nucleons to nuclei}
\label{sec:nuc}
In this section we provide further details on the relation between differential cross section for DM-nucleus scattering and the underlying nuclear charges and currents (see Eq.~(\ref{eq:matrix_element2})).~We also provide additional information on the DM response functions upon which such cross section depends.~The nuclear charges and currents appearing in Eq.~(\ref{eq:matrix_element2}) are defined as follows
\begin{align}
\label{eq:ncv}
    S &= \sum_{i=1}^A e^{-i\vec{q}\cdot \vec{x}_i}\nonumber \\
    T &= \sum_{i=1}^A 
\frac{1}{2m_N}\left\{i\overleftarrow{\nabla}_i\cdot\vec{\sigma}_ie^{-i\vec{q}
\cdot \vec{x}_i} - i e^{-i\vec{q}\cdot 
\vec{x}_i}\vec{\sigma}_i\cdot\overrightarrow{\nabla}_i\right\}\nonumber\\
    \vec{P} &= \sum_{i=1}^A \vec{\sigma}_ie^{-i\vec{q}\cdot \vec{x}_i}\nonumber\\
    \vec{Q} &= \sum_{i=1}^A 
\frac{1}{2m_N}\left\{i\overleftarrow{\nabla}_ie^{-i\vec{q}\cdot \vec{x}_i} - 
i e^{-i\vec{q}\cdot \vec{x}_i}\overrightarrow{\nabla}_i\right\}\nonumber\\
    \vec{R} &= \sum_{i=1}^A 
\frac{1}{2m_N}\left\{\overleftarrow{\nabla}_i\times\vec{\sigma}_i e^{-i\vec{q}
\cdot \vec{x}_i} + e^{-i \vec{q}\cdot 
\vec{x}_i}\vec{\sigma}_i\times\overrightarrow{\nabla}_i\right\}\,
    \end{align} 
where $\vec{x}_i$ is the position vector of the $i$-th nucleon bound in the nucleus, the gradient operator $\overrightarrow{\nabla}_i$ acts on the coordinates of the $\vec{x}_i$ vector, and $\vec{{\sigma}}_i$ is a vector of Pauli matrices acting on the spin space of the $i$-th constituent nucleon.~The above nuclear charges and currents arise from the expansion in spherical harmonics of the nuclear potential that we obtained from $\mathcal{M}^{(0)}$ and $\mathcal{M}^{(1)}$, and involve a summation over all constituent nucleons (here $A$ is the atomic mass number).~The gradient operators in the above expressions provide a position-space representation for the individual nucleon velocities.~See~\cite{Fitzpatrick:2012ix} for further details on the derivation of Eq.~(\ref{eq:ncv}).~We conclude this appendix by providing the explicit relation between the DM response functions that we found in~Eq.~(\ref{eq:DMresponses}), and the $\ell$-coefficients that appear in Eq.~(\ref{eq:matrix_element2}):
\begin{align}
R^{\tau\tau'}_{M} &= \frac{1}{3}\sum_{s,s'} 
\ell_0^{\tau}\ell_0^{\tau' *} \nonumber\\
\qpmn{2}R^{\tau\tau'}_{\Phi''} &= 
\frac{1}{3}\sum_{s,s'}\qmn\cdot\vec{\ell}_{E}^{\tau}\,
\qmn\cdot\vec{\ell}_E^{\tau' *}\nonumber\\
\qpmn{2}R^{\tau\tau'}_{\Phi''M} &= \frac{1}{3}\sum_{s,s'} 
\tqmn\cdot\Re\left[\vec{\ell}_{E}^{\tau}\ell_0^{\tau' *}\right]\nonumber\\
\qpmn{2}R^{\tau\tau'}_{\tilde{\Phi}'} &= 
\frac{1}{3}\sum_{s,s'}\frac{1}{2}\left(\qpmn{2}\vec{\ell}_{E}^{
\tau}\cdot\vec{\ell}_E^{\tau' *} - 
\qmn\cdot\vec{\ell}_{E}^{\tau}\,\qmn\cdot\vec{\ell}_E^{
\tau' *}\right) \nonumber\\
R^{\tau\tau'}_{\Sigma''} &= 
\frac{1}{3}\sum_{s,s'}\hat{q}\cdot\vec{\ell}_5^{\tau}\,\hat
{q}\cdot\vec{\ell}_5^{\tau' *} \nonumber\\
R^{\tau\tau'}_{\Sigma'} &= 
\frac{1}{3}\sum_{s,s'}\frac{1}{2}\left(\vec{\ell}_5^{\tau}
\cdot\vec{\ell}_5^{\tau' *} - 
\hat{q}\cdot\vec{\ell}_5^{\tau}\,\hat{q}\cdot\vec{\ell}_5^{
\tau' *}\right) \nonumber\\
\qpmn{2}R^{\tau\tau'}_{\Delta} &= 
\frac{1}{3}\sum_{s,s'}\frac{1}{2}\left(\qpmn{2}\vec{\ell}_{M}^{
\tau}\cdot\vec{\ell}_M^{\tau' *} - 
\qmn\cdot\vec{\ell}_{M}^{\tau}\,\qmn\cdot\vec{\ell}_M^{
\tau' *}\right) \nonumber\\
\qpmn{2}R^{\tau\tau'}_{\Delta\Sigma'} &= \frac{1}{3}\sum_{s,s'} 
\qmn\cdot\Re\left[i\vec{\ell}_{M}^{\tau}\times\vec{\ell}_{5
}^{\tau' *} \right]\,.
\label{eq:Rl2}
\end{align}
In the above expressions, angle brackets denote an expectation value between initial and final DM spin states, and $\hat{q}$ is a unit vector pointing in the direction of the momentum transfer.

\section{Summation rules for spin 1 DM}
In this appendix we provide a list of spin summation rules (some involving the $\sy$ operator) that apply to spin 1 DM:
\label{sec:app_summation_rules}
\begin{align}
&\frac{1}{3}\sum_{s,s'}\vec{A}\cdot\sX = 0 \nonumber \\ 
&\frac{1}{3}\sum_{s,s'}\left(\sX\right)^2 = 2 \nonumber\\ 
&\frac{1}{3}\sum_{s,s'}\left(\vec{A}\cdot\sX\right)\left(\vec{B}\cdot\sX\right) = 
\frac{2}{3}\vec{A}\cdot \vec{B}\nonumber\\ 
&\frac{1}{3}\sum_{s,s'}
\left(\vec{A}\times\sX\right)\cdot\left(\vec{B}\times\sX\right) = \frac{4}{3}\vec{A}\cdot \vec{B}\nonumber\\ 
&\frac{1}{3}\sum_{s's}\left(\vec{A}\cdot\sy^{s's}\cdot \vec{B}\right)\delta^{ss'} = 
\frac{1}{3}\vec{A}\cdot \vec{B} \nonumber\\ 
&\frac{1}{3}\sum_{s's}\vec{A}\cdot\sy^{s's}\cdot \sX = 0\nonumber\\ 
&\frac{1}{3}\sum_{s's}\left(\vec{A}\cdot\sy^{s's}\cdot \vec{B}\right)\left(\sX\cdot 
\vec{C}\right) = 0\nonumber\\ 
&\frac{1}{3}\sum_{s's}\left(\vec{A}\cdot\sy^{s's}\right)\cdot\left(\vec{B}\cdot\sy^{s's}\right) 
= \frac{2}{3}\vec{A}\cdot \vec{B}\nonumber\\ 
&\frac{1}{3}\sum_{s's}|\vec{A}\cdot\sy^{s's}\cdot \vec{B}|^2 = \frac{1}{6}|\vec{A}\cdot \vec{B}|^2 + 
\frac{1}{6}|\vec{A}|^2|\vec{B}|^2  \,,
\end{align}
where $\vec{A}$, $\vec{B}$ and $\vec{C}$ are arbitrary three-dimensional vectors.~These expressions are useful when evaluating Eq.~(\ref{eq:Rl2}).

\providecommand{\href}[2]{#2}\begingroup\raggedright\endgroup
\end{document}